\documentclass[journal=jacsat,manuscript=article]{achemso}
\usepackage{achemso}
\setkeys{acs}{maxauthors = 0}
\usepackage[version=3]{mhchem}
\usepackage[english]{babel}
\usepackage{filecontents}
\usepackage{makecell}
\usepackage{graphicx}
\usepackage{courier}
\usepackage{amssymb,amsmath,amsthm,mathrsfs,comment,afterpage,accents}
\usepackage{mathtools}
\usepackage{braket}
\usepackage{mathrsfs}
\usepackage{courier}
\usepackage{color}
\usepackage{subdepth}
\usepackage{tablefootnote}
\usepackage{hyperref}
\usepackage[capitalize]{cleveref}
\usepackage[linesnumbered,ruled,vlined]{algorithm2e}
\setkeys{acs}{articletitle = true}
\usepackage{xr}
\usepackage{xc}
\externalcitedocument[]{supporting_information}
\makeatletter
\SectionNumbersOn
\def\@dotsep{4.5}
\usepackage{cancel}
\makeatother
\usepackage{dcolumn}
\usepackage{bm}
\newcommand\mat\mathbf

\newcommand{\insertnew}[1]{{\textcolor{black} {#1}}}
\newcommand{\insertrev}[1]{{\textcolor{black} {#1}}}
\newcommand{\revrep}[2]{{\textcolor{red}{}}{\textcolor{black}{#2}}}

\title{
Systematically Improvable Tensor Hypercontraction:
Interpolative Separable Density-Fitting for Molecules
Applied to
Exact Exchange, Second- and Third-Order M{\o}ller-Plesset Perturbation Theory
}

\author{Joonho Lee}
\email{linusjoonho@gmail.com}
\affiliation{
Department of Chemistry, University of California, Berkeley, California 94720, USA
Chemical Sciences Division, Lawrence Berkeley National Laboratory, Berkeley, California 94720, USA
}
\author{Lin Lin}
\affiliation{
Department of Mathematics, University of California, Berkeley, California 94720, USA
Computational Research Division, Lawrence Berkeley National Laboratory, Berkeley, California 94720, USA
}
\author{Martin Head-Gordon}
\email{mhg@cchem.berkeley.edu}
\affiliation{
Department of Chemistry, University of California, Berkeley, California 94720, USA
Chemical Sciences Division, Lawrence Berkeley National Laboratory, Berkeley, California 94720, USA
}
\begin{document}
\newpage
\maketitle
\begin{abstract}
We present a systematically improvable tensor hypercontraction (THC) factorization
based on interpolative separable density fitting (ISDF).
We illustrate algorithmic details to achieve this within the framework of Becke's atom-centered quadrature grid.
A single ISDF parameter $c_\text{ISDF}$ controls the tradeoff between accuracy and cost.
In particular, $c_\text{ISDF}$ sets the number of interpolation points used in THC, $N_\text{IP} = c_\text{ISDF}\times N_\text{X}$ with $N_\text{X}$ being the number of auxiliary basis functions.
In conjunction with the resolution-of-the-identity (RI) technique,
we develop and investigate the THC-RI algorithms for
cubic-scaling exact exchange for Hartree-Fock and range-separated hybrids (e.g., $\omega$B97X-V) and quartic-scaling second- and third-order M{\o}ller-Plesset theory (MP2 and MP3).
These algorithms were evaluated over the W4-11 thermochemistry (atomization energy) set and A24 non-covalent interaction benchmark set
with standard Dunning basis sets (cc-pVDZ, cc-pVTZ, aug-cc-pVDZ, and aug-cc-pVTZ).
We demonstrate the convergence of THC-RI algorithms to numerically exact RI results using ISDF points.
Based on these, we make recommendations on $c_\text{ISDF}$ for each basis set and method.
\insertrev{We also demonstrate the utility of THC-RI exact exchange and MP2 for larger systems such as water clusters and \ce{C20}.}
We stress that more challenges await in obtaining accurate and numerically stable THC factorization for \insertrev{wavefunction amplitudes as well as} the space spanned by virtual orbitals in large basis sets and
implementing sparsity-aware THC-RI algorithms.
\end{abstract}
\newpage
\section{Introduction} \label{sec:intro}
\insertnew{
The introduction of a single particle basis of atomic orbitals (AOs), $\{\omega_\mu(\mathbf r)\}$, and the presence
of $\hat{V}_\text{ee} = \sum_{i<j} r_{ij}^{-1}$ in the electronic Hamiltonian leads to the 
4-leg two-electron repulsion integral (ERI) tensor,
\begin{equation}
\left(\mu\nu|\lambda\sigma\right)
= 
\int \mathrm d \mathbf r_1\int \mathrm d \mathbf r_2
\frac{
\omega_\mu(\mathbf r_1) \omega_\nu(\mathbf r_1)
\omega_\lambda(\mathbf r_2) \omega_\sigma(\mathbf r_2)
}{|\mathbf r_1 - \mathbf r_2|}.
\label{eq:2e}
\end{equation}
as a central quantity in computational quantum chemistry\cite{Pople1978,Rys1983,Obara1986, HeadGordon1988, gill1989efficient, Gill1991,Gill1991a}.
Spatial locality of the $N_\text{AO}$ AOs means that only a quadratic number of
the $\mathcal O(N_\text{AO}^4)$ ERIs are numerically significant, which 
has enabled efficient AO-driven methods for Hartree-Fock (HF) and density functional theory \cite{almlof1982principles,haser1989improvements} as well as some post-HF methods \cite{haser1993moller,koch1994direct}.
Further screening and/or collectivization of long-range quantities permits development of linear-scaling methods as well \cite{White1994, White1996, Ochsenfeld1998, ochsenfeld2007linear, Doser2009}.
However such methods involve large prefactors because of the cost of 
handling the significant 4-leg ERIs.
}

\insertnew{Therefore there has been much interest in methods
that entirely avoid 4-leg ERI evaluation, and instead
use factorization in terms of lower rank quantities.
A graphical representation of the approach reviewed below is provided in \cref{fig:thc}.
}
The most widely used integral
factorization is 
undoubtedly 
the resolution-of-the-identity (RI) (also called density-fitting (DF)) technique \cite{Baerends1973, Whitten1973,Jafri1974}.
The key idea in the RI factorization is that 
one defines an auxiliary basis set $\{\omega_P\}$ which approximately represents electron density 
of system.
This approximation can be made ``optimal'' via least-squares fit within
some metric (most commonly Coulombic)\cite{Vahtras1993}.
Such fitting factorizes the 4-leg tensor in \cref{eq:2e} into 3-leg and 2-leg tensors,
\begin{equation}
(\mu\nu|\lambda\sigma)
\approx \sum_{PQ}C_{\mu\nu}^P (P|Q) C_{\lambda\sigma}^Q
\end{equation}
\begin{equation}
C_{\mu\nu}^P = \sum_Q
(\mu\nu|Q) \left[(Q|P)\right]^{-1}
\end{equation}
and
thereby often greatly simplifies subsequent manipulations of the ERI tensor.
The size of the auxiliary basis set, $N_X$, used in RI to fit
the electron density is roughly 3-4 times larger than the original basis set
to achieve the accuracy of 50-60 $\mu E_h$ per atom \cite{Eichkorn1995, Eichkorn1997}.
We note that the theoretical understanding of RI was first provided by Dunlap \cite{Dunlap1979,Dunlap1983, Dunlap2000, Dunlap2000a}.
The RI technique has been combined with
some local fitting strategies \cite{werner2003fast, Polly2004, Sodt2006, Merlot2013, Hollman2014, Manzer2015a} to achieve linear scaling.
From a similar perspective, Beebe and Linderberg applied Cholesky decomposition (CD) to the two-electron integrals
which in turn achieves more or less the same speedup and compactness as those of the RI factorization \cite{Beebe1977}.
It is now routine to apply the RI approximation and related techniques
to HF theory \cite{fruchtl1997implementation, manzer2015fast}, M{\o}ller-Plesset (MP) perturbation theory \cite{Feyereisen1993, Bernholdt1996}, coupled-cluster (CC) theory \cite{Rendell1994, Epifanovsky2013} etc.

From a somewhat different perspective,
Friesner applied a pseudospectral (PS) method to approximate the ERI tensor \cite{Friesner1985,Friesner1986,Friesner1987}. 
In PS, we define a secondary basis set as points in 3-space.
The key idea is to perform one of the two integrals in \cref{eq:2e}
analytically and the rest is evaluated numerically,
which avoids the Coulomb singularity and provides a numerically stable algorithm.
\insertnew{
Namely, 
\begin{align}
(\mu\nu|\lambda\sigma) &\approx
\sum_g \lambda_g 
\omega_\mu(\mathbf r_g)
\omega_\nu(\mathbf r_g)
V_{\lambda\sigma}(\mathbf r_g)\\
V_{\lambda\sigma} (\mathbf r_g)
&= \int\mathrm d \mathbf r \frac{\omega_\lambda(\mathbf r)\omega_\sigma(\mathbf r)}{|\mathbf r - \mathbf r_g|}
\end{align}
where $\{r_g\}$ is a set of 3-space points and $\{\lambda_g\}$ is the weight associated with each grid point.}
This simple idea was successfully applied to HF achieving 
an overall cubic scaling algorithm, which is lower-scaling than the conventional RI-driven HF theory (quartic scaling) \cite{Ringnalda1990}.
This was further extended to full configuration interaction \cite{Martinez1992}, 
configuration interaction with doubles \cite{Martinez1993}, 
and the second- and the third-order MP (MP2 and MP3) \cite{Martinez1994} by Mart{\'i}nez and Carter.
The major drawback of those PS methods 
is the lack of generality. That is,
to achieve both efficiency and accuracy
quoted in refs. \citenum{Ringnalda1990, Won1991, Greeley1994} computational algorithms need to be optimized for specific basis sets.
It is noteworthy that Neese and co-workers'  chain-of-sphere approximation for exact exchange (COSX) is particularly
promising in terms of general applicability and simple implementation \cite{Neese2009}.

A more flexible integral factorization was proposed by Mart{\'i}nez and co-workers,
termed tensor hypercontraction (THC) \cite{Hohenstein2012, Parrish2012, Hohenstein2012a, Parrish2013a, Hohenstein2013a, Hohenstein2013, Parrish2013, Parrish2014, KokkilaSchumacher2015, Song2016, Song2017, Song2017a}.
It is more flexible because the 4-leg tensor in \cref{eq:2e}
is factorized into multiplications among five matrices:\insertnew{
\begin{equation}
\left(
\mu\nu|\lambda\sigma
\right)
\approx
\sum_{KL}
\omega_\mu(\mathbf r_K) \omega_\nu(\mathbf r_K)
M_{KL}
\omega_\lambda (\mathbf r_L) \omega_\sigma(\mathbf r_L)
\label{eq:thc}
\end{equation}
where $\{\mathbf r_K\}$ are chosen grid points and $M_{KL}$ will be discussed in detail later.} This factorization therefore requires only {\it quadratic} storage which is lower than the cubic storage generally needed in RI.
Initial attempts to obtain such flexible \insertrev{and in principle systematically improvable} factorization based on CANDECOMP/PARAFAC (PF)
were found to be too inefficient to be competitive with conventional RI algorithms \cite{Hohenstein2012}.
\insertrev{Nevertheless, there have been several attempts to directly obtain the PF-THC factorization. Benedikt and co-workers studied CC with doubles with THC factorization obtained via $\mathcal O(N^5)$ alternating least-squares algorithm \cite{benedikt2013tensor}. Hummel and co-workers obtained a PF-THC factorization based on $\mathcal O(N^4)$ alternating least-squares and the resulting factorization was applied to planewave CC with singles and doubles (CCSD) \cite{hummel2017low}. A comparison of different strategies ($\mathcal O(N^{4-5})$ algorithms) for obtaining PF-THC factorization in the context of gaussian orbital based CCSD was done by Schutski and co-workers \cite{schutski2017tensor}.}

A different strategy using least-squares fitting, namely least-squares THC (LS-THC), 
was found to be more efficient, \insertrev{in principle systematically improvable,} and \insertrev{yet still} accurate \cite{Parrish2012}.
\insertrev{Unlike the aforementioned approach, obtaining the THC factorization in LS-THC scales cubically with system size when sparsity is considered.}
Some intermediates in LS-THC have a strong resemblance with PS approaches but overall it provides
more convenient factorization for subsequent calculations.
Combined with the RI technique, the resulting LS-THC-RI has been
applied to MP2 \cite{Hohenstein2012}, MP3 \cite{Hohenstein2012}, CCSD \cite{Hohenstein2012a}, second-order approximate CC with singles and doubles for ground state \cite{Hohenstein2013a} and excited states \cite{Hohenstein2013}, 
second-order complete active-space perturbation theory (CASPT2) \cite{Song2018},
particle-particle random phase approximation (ppRPA)\cite{Shenvi2014}, and direct random phase approximation (dRPA) \cite{Duchemin2019} for finite-sized systems (i.e., molecules).
Although LS-THC-RI is powerful on its own,
to realize its full potential 
some optimization of quadrature points and weights was needed for specific basis sets \cite{KokkilaSchumacher2015}.

In our opinion, 
\insertrev{the major reason behind somewhat uncommon usage of LS-THC-RI by research groups other than its developers
is its lack of generality in generating a compact set of grid points that works for all chemical elements and any basis set. } Recently, Lu and Ying recognized that LS-THC is a form of a low-rank approximation to the
density represented in 3-space \cite{Lu2015}. 
This unique perspective allowed for a systematic
way to pick a set of 3-space points used in THC based on column-pivoted QR (QRCP) decomposition.
This was termed interpolative separable density fitting (ISDF). Those points are
not quadrature points unlike in the original formulation of THC.
Rather, they are {\it interpolation} points which therefore are not associated with any quadrature weights.
Subsequently, the centroidal Voronoi tessellation (CVT) technique was developed
to remove the expensive QRCP step and reduce the overall cost \insertrev{to $\mathcal O(N^2)$} for obtaining interpolation points \cite{Dong2018}.
The ISDF approach was then applied to HF exact exchange (HF-K) \cite{Hu2017}, dRPA \cite{Lu2017}, the Bethe-Salpeter equation \cite{Hu2018}, 
phonon calculations in density functional perturbation theory \cite{lin2017adaptively},
and auxiliary-field quantum Monte Carlo \cite{malone2018overcoming}
for systems under periodic boundary conditions (PBCs).

Given the utility of the ISDF point selection shown for systems under PBCs,
it is valuable to apply this technique to molecular simulations 
where the original \insertrev{LS-}THC techniques currently lack full generality \insertrev{in choosing a compact set of 3-space points}. 
There are some technical challenges: (1) a fully numerical evaluation of Coulomb integrals
is nearly impossible for finite-sized systems due to the Coulomb singularity in contrast to
those for infinite systems where one can use fast Fourier transform (FFT) and (2) we use 
atom-centered quadrature grids \cite{Becke1988} unlike uniform grids used for PBC simulations. 
We will show that (1) has already been solved in LS-THC-RI
and for (2) we will present a modified \insertrev{$\mathcal O(N)$} CVT algorithm that is well-suited for atom-centered grids. 
Furthermore, we will show that performing least-squares fits for each molecular orbital (MO) block in the MO transformed integrals 
is much more accurate than fitting the atomic orbital (AO) integrals, which was \revrep{also}{previously} noted in refs. \citenum{Parrish2012,Parrish2014} \insertrev{by Mart{\'i}nez's group}.
\insertrev{In passing, we note that our CVT procedure has the same goal as the radial discrete value representation (R-DVR) developed by Mart{\'i}nez and co-workers \cite{Parrish2013a}. R-DVR achieves generality just like CVT as it, in principle, can work with all chemical elements and basis sets without empirical fitting. However, R-DVR was shown to be less compact than hand-optimized grids for water clusters for similar accuracy\cite{KokkilaSchumacher2015}. On the other hand, the grid points obtained from CVT were found to be as compact and accurate as the hand-optimized ones as discussed later in this work. Therefore, the CVT procedure examined here achieves both compactness and generality.}

\insertrev{In addition to the atom-centered CVT procedure, we also test the accuracy of the factorization of the MP1 $T$-amplitudes when evaluating MP2 and MP3 correlation energies. The factorization of $T$-amplitude has already been used in THC-CCSD \cite{Hohenstein2012a,benedikt2013tensor,Shenvi2014,schutski2017tensor}.
Such factorization can speed up THC-RI-MP3 by a factor of 49 compared to the original algorithm \cite{Hohenstein2012} by removing the double Laplace transform associated with the algorithm. Our goal is to examine the accuracy of this factorization when using CVT grid points and the AO basis functions for least-squares fits over a wide range of small molecules. }
This paper is organized as follows:
we will (1) review the essence of THC factorization and ISDF point selection,
(2) present a modified CVT algorithm for atom-centered grids,
(3) \revrep{present}{review} THC algorithms for exact exchange, MP2, and MP3,
(4) present careful benchmarks of the accuracy of these algorithms
over the W4-11 \cite{Karton2011} and A24 \cite{Rezac2013} sets (widely used thermochemistry and non-covalent interaction benchmarks, respectively) \insertrev{as well as water clusters and \ce{C20}}.

\section{Theory}
\subsection{Notation}
In Table \ref{tab:notation}, we summarize the notation used in this work.
\begin{table}[h]
\begin{tabular}{c|c}
Notation                                & Description     \\\hline
$\mu$, $\nu$, $\lambda$, $\sigma$, etc. & indices for atomic orbitals (AOs) \\\hline
$i$, $j$, $k$, $l$, etc. & indices for occupied molecular orbitals (MOs) \\\hline
$a$, $b$, $c$, $d$, etc. & indices for virtual molecular orbitals (MOs) \\\hline
$P$,$Q$,$R$,$S$                                        &     indices for resolution-of-the-identity (RI) basis functions\\\hline
$K$,$L$,$M$,$N$ &  indices for interpolation vectors\\\hline
$N_\text{AO}$ & the number of AOs\\\hline
$N_\text{X}$ & the number of RI basis functions\\\hline
$N_\text{SP}$ & the number of {\it significant} shell-pairs \\\hline
$N_\text{IP}$ & the number of interpolation vectors \\\hline
$n_\text{occ}$ & the number of occupied orbitals \\\hline
$n_\text{vir}$ & the number of virtual orbitals \\\hline
$n_t$ & the number of Laplace quadrature points \\\hline
$\omega_\mu(\mathbf r)$ & $\mu$-th AO basis function at $\mathbf r$\\ \hline
$\omega_P(\mathbf r)$ & $P$-th auxiliary (RI) basis function at $\mathbf r$ \\ \hline
$\phi_p(\mathbf r)$ & $p$-th molecular orbital (MO) at $\mathbf r$ \\ \hline
$\xi_K (\mathbf r)$ & $K$-th interpolation vector at $\mathbf r$ \\\hline
\end{tabular}
\caption {Notation used in this work.}
\label{tab:notation}
\end{table}
We assume the spin-orbital basis unless specified otherwise.
We define molecular orbitals (MOs) with MO coefficient matrix, $\mathbf c$, 
\begin{equation}
\phi_p (\mathbf r) = \sum_\mu c_{\mu p} \omega_\mu (\mathbf r)
\label{eq:ao2mo}
\end{equation}
For later use, we define {\it imaginary time-dependent} MOs:
\begin{align}
\phi_{i}(t) &= \phi_{i} e^{-t \epsilon_i}\\
\phi_{a}(t) &= \phi_{a} e^{t \epsilon_a}
\end{align}
where
$\epsilon_p$ denotes the orbital energy of the $p$-th MO.
These time-dependent MOs naturally arise
when applying the Laplace transformation to orbital energy denominators as shown later.
\subsection{Review of Tensor Hypercontraction}
We review tensor hypercontraction (THC) developed by Mart{\'i}nez and co-workers \cite{Hohenstein2012, Parrish2012, Hohenstein2012a, Parrish2013a, Hohenstein2013a, Hohenstein2013, Parrish2013, Parrish2014, KokkilaSchumacher2015, Song2016, Song2017, Song2017a}. 
In resolution-of-the-identity (RI) or density-fitting (DF), 
\insertnew{product of AOs are represented in an auxiliary basis by the following decomposition of}
a 3-leg density tensor $Z_{\mu\nu}(\mathbf r)$:
\begin{equation}
Z_{\mu\nu}(\mathbf r) = \omega_\mu(\mathbf r) \omega_\nu (\mathbf r)
\simeq
 \sum_{P}
C_{\mu\nu}^{P} \omega_{P}(\mathbf r) 
\label{eq:Zmn}
\end{equation}
where
$\omega_\mu(\mathbf r)$ is the AO basis function on a grid point $\mathbf r$,
$C_{\mu\nu}^{{P}}$ is the fit coefficient, and $\omega_{ P}(\mathbf r)$ is the auxiliary function.
The use of the term RI is ironic in that the product-separable structure in $Z_{\mu\nu}(\mathbf r)$ disappears with it.
Namely, RI introduces an inseparable 3-leg tensor as the 3-leg fit coefficient $C_{\mu\nu}^{ P}$.
The upshot of THC factorization is that the 3-leg fit coefficient in RI is avoided by use of
an approximate separable form involving a grid of points:
\begin{equation}
Z_{\mu\nu}(\mathbf r)
\simeq
 \sum_{K}
C_{\mu\nu}^{K}
\xi_{K}(\mathbf r) 
=
 \sum_{K}
\omega_\mu^{K}
\omega_\nu^{K}
\xi_{K}(\mathbf r) 
\label{eq:fit}
\end{equation}
where we write
\begin{equation}
\omega_\mu^{K}
\equiv
\omega_\mu(\mathbf r_{K})
\end{equation}
and
$\xi_K(\mathbf r)$
is a fit function (or interpolation vector) that will be specified later.
\begin{figure}[h!]
\includegraphics[scale=0.55]{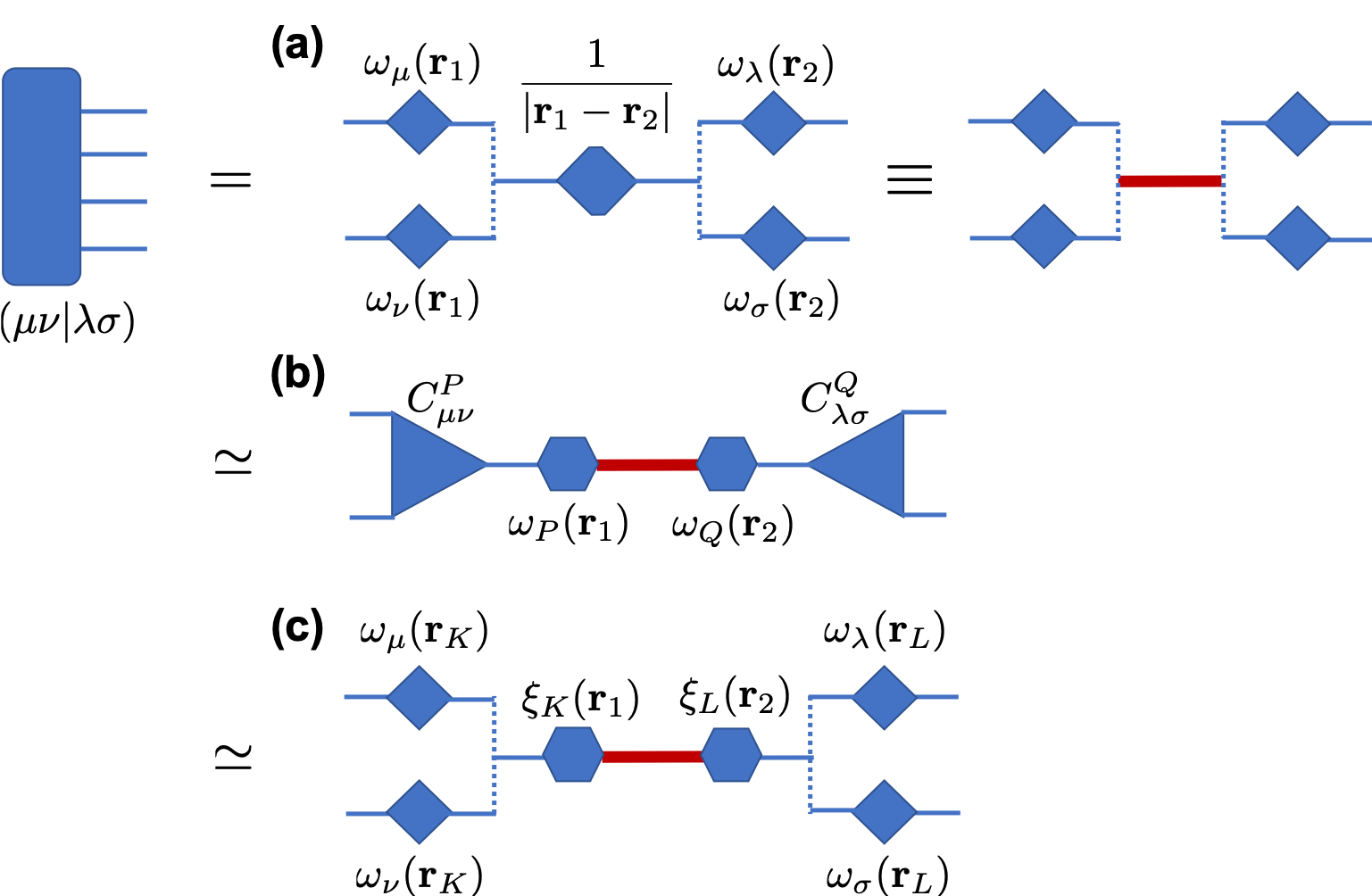}
\caption{
Graphical representation of
density tensor
(a) represented exactly,
(b) represented using RI,
and
(c) represented using THC.
The solid blue line is a contraction over shared indices between two tensors
while the dotted blue line indicates element-wise product over connected indices (i.e., Hadamard product).
The thick red bar represents a contraction (in real-space) with the Coulomb operator, $\frac{1}{|\mathbf r_1 - \mathbf r_2|}$.
}
\label{fig:thc}
\end{figure}

The key differences between RI and THC in representing the 4-leg integral tensor, $(\mu\nu|\lambda\sigma)$, can be also understood graphically as described in Figure \ref{fig:thc}.
The RI representation evidently removes the separable structure in $\mathbf Z$ whereas
the THC factorization recovers such a separable form.
This separable structure is the key in reducing the cost for subsequent calculations.
Achieving the THC factorization takes two crucial steps:
(1) choosing a set of points $\{\mathbf r_{K}\}$ in 3-space and (2) determining the fit functions $\{\xi_{K}(\mathbf r)\}$.
\insertnew{Their number, $N_\text{IP}$, must be as small as possible for efficiency, but as large as necessary to achieve usable accuracy.}
For (1), we will follow the strategy introduced by Lu and Ying \cite{Lu2015}. Namely, we will use the interpolative separable density fitting (ISDF) technique to determine a set of {\it interpolation points}. Then, for (2), we will follow the least-squares THC (LS-THC) technique by Mart{\'i}nez and co-workers \cite{Parrish2012}.
For the purpose of this paper, we will refer to LS-THC simply as THC.

\subsection{Selecting Interpolation Points}
When choosing interpolation points $\{\mathbf {r}_{K}\}$,
Mart{\'i}nez and co-workers have tried to prune the Becke quadrature grid widely used in density functional theory \cite{Becke1988}.
Some of their efforts have focused on further optimizing the grid representation \cite{Parrish2013a, KokkilaSchumacher2015} and
special optimization procedures
were performed for each basis set \cite{KokkilaSchumacher2015}.
A simpler and automatic interpolation point generation is possible as suggested in ISDF.
We will describe how one can adapt the ISDF procedure developed in refs. \citenum{Lu2015, Dong2018} to choose a subset of Becke quadrature points. This procedure is almost automatic and also 
systematically improvable with a single tunable parameter.

The important observation made in ref. \citenum{Lu2015} is that \cref{eq:fit} is in fact a form of a low-rank approximation
\insertnew{when evaluated on a finite grid of points, $N_g$}.
If one looks at $\mathbf Z$ as a $N_g$-by-$N_\text{AO}^2$ matrix, 
a singular value decomposition (SVD) of $\mathbf Z$ would have at most min($N_g$, $N_\text{AO}^2$) number of non-zero singular values.
In most practical applications, $N_\text{AO}^2$ is greater than $N_g$.
The goal of ISDF is then to pick a subset of $N_g$ rows of $\mathbf Z$ that
closely approximates the entire matrix $\mathbf Z$. Such a low-rank approximation is called
interpolative decomposition (ID) where the goal is to compress a given tensor while preserving the column or row structure.
A common way to accomplish ID is to use the QR decomposition with column-pivoting (QRCP) as was done in ref. \citenum{Lu2015}. 
Namely,
\begin{equation}
\mathbf Z^T \mathbf \Pi = \mathbf Q \mathbf R
\end{equation}
where $\mathbf \Pi$ is the permutation matrix from the QRCP procedure, $\mathbf Q$ is an orthogonal matrix, and $\mathbf R$ is a upper triangular matrix.
Each column of $\mathbf \Pi$ contains a single 1 and $(N_g-1)$ 0's. It permutes the columns of $\mathbf Z^T$ such that 
the diagonal entires of $\mathbf R$ are non-increasing. In other words, the $\mathbf \Pi$ matrix orders the columns of $\mathbf Z^T$ based on their relative importance. Therefore, this automatically produces an ``optimal'' subset of $N_g$ grid points that are used as interpolation points in ISDF.
The cost of this QRCP is quartic if performed exactly and can be reduced to cubic if randomized algorithms are employed \cite{Lu2015, Hu2017}.

Rather than employing randomized QRCP, a further cost reduction in choosing interpolation points is possible using the centroidal Voronoi tessellation (CVT) procedure proposed by Lin and co-workers \cite{Dong2018}.
We will describe a modified version of the CVT procedure so that it is well-suited for Becke's quadrature grid \cite{Becke1988}.
Unlike uniform grids, the Becke grid points carry quadrature weights which also encode some geometric information of systems through interatomic quadrature weights, $\lambda(\mathbf r)$. In this sense, a better quantity to compress is 
\begin{equation}
\lambda(\mathbf r) Z_{\mu\nu}(\mathbf r) 
\simeq
 \sum_{K}
C_{\mu\nu}^{K} \lambda(\mathbf r_{K})
\xi_{K}(\mathbf r) .
\label{eq:wZ}
\end{equation}
The quadrature weights are necessary to account for non-uniformity of the Becke grid and
it was indeed found to be much more accurate than carrying out interpolation without the weights.
We note that
one can trivially compute an overlap matrix element on the Becke grid:
\begin{equation}
S_{\mu\nu} = 
\sum_\mathbf{r} 
\sum_{K}
C_{\mu\nu}^{K}
\lambda(\mathbf r_{K})
\xi_{K}(\mathbf r)
\end{equation}
which is a useful quantity for debugging purposes.

In Algorithm \ref{algo:cvt}, we outline the CVT procedure adopted for the Becke grid.
In essence, we start from an initial set of centroids, $\mathbf{c}^{(0)}$, and update
these points based on the neighbors of these points as well as the weights associated with them.
The resulting centroids are our interpolation points.
Interested readers are referred to ref. \citenum{Dong2018}. It is worthwhile to mention two modifications
we made to generate the interpolation points
as a subset of the Becke grid.

First, we loop over individual atoms to generate atom-specific interpolation points.
This is necessary to maximally exploit the locality of Gaussian AO basis sets and atom-centered grids. 
Furthermore, in Algorithm \ref{algo:cvt} the distance between centroids and grid points needs to be computed
every iteration and this becomes quite computationally demanding.
If one were to do this for all atoms at once, this step scales quadratically with system size and carries a non-negligible prefactor.
Looping over individual atoms naturally yields an algorithm that scales linearly with system size.

Second, we use the Becke quadrature weights in the weighted CVT procedure.
The Becke quadrature weights reflect geometric information around a given grid point through the so-called Becke partition function \cite{Becke1988}.
Namely, the weight of a grid point depends on the location of all atoms and associated grid points.
Lin and co-workers suggested that electron density $\rho(\mathbf r)$ can be used for the weight function in CVT.
The use of $\rho(\mathbf r)$ in the CVT procedure will complicate derivative theories since the position of those 3-space points will depend on the electronic density.
As such, we considered other alternatives and we found that the Becke quadrature weights are simple to use and yield interpolation points that result in accurate compression for later uses. This will not complicate derivative theories either.
This weighted CVT procedure is used throughout this paper.

\begin{algorithm}[H]
\label{algo:cvt}
\SetAlgoLined
\For{$A=1$ \KwTo $N$\tcp*{Loop over atoms.}}
{
Generate $n_g$ grid points, $\{\mathbf r_A\}$, and their weights, $\lambda(\mathbf r)$ for $\mathbf r \in \{\mathbf r_A\}$\; 
Compute $\omega(\mathbf r)$ for $\mathbf r \in \{\mathbf r_A\}$\;
Select $n_\mu$ points, $\mathbf{c}^{(0)}$, randomly from $\{\mathbf r_A\}$\;
Call the K-means subroutine described in the Algorithm I in ref. \citenum{Dong2018} with weight function $\lambda(\mathbf r)$\;
}
\caption{Weighted K-Means Algorithm For the Becke Grid. 
$n_g$ and $n_\mu$ denote the number of quadrature points and the number of
interpolation points (user-specified) for a given atom, respectively.
}
\end{algorithm}

\subsection{Determination of Interpolation Vectors}

Once we have a set of interpolation points $\{\mathbf r_{K}\}$, the hitherto undefined interpolation vectors 
required in \cref{eq:fit}
can be obtained via the following least-squares expression,
\begin{equation}
\xi_{K}(\mathbf r) = 
\sum_{L}
\tilde{S}^+_{KL}
C_{\mu\nu}^L \lambda(\mathbf r_L)
Z_{\mu\nu}(\mathbf r) \lambda(\mathbf r)
\label{eq:xi}
\end{equation} 
where 
$\mathbf Z$ and $\mathbf C$ are defined in \cref{eq:Zmn} and \cref{eq:fit}, respectively, and 
the metric in 3-space is defined as
\begin{equation}
\tilde {S}_{K L} 
= 
\sum_{\mu\nu}
\omega_\mu^K
\omega_\nu^K
\lambda(\mathbf r_K)
\omega_\mu^L
\omega_\nu^L
\lambda(\mathbf r_L)
\label{eq:metric}
\end{equation}
and $\tilde{\mathbf S}^{+}$ is the pseudoinverse of $\tilde{\mathbf S}^{+}$.
$\tilde{\mathbf S}$ is generally ill-conditioned \cite{Song2017a} and the threshold of $10^{-12}$ in the drop tolerance for evaluating $\tilde{\mathbf S}^{+}$ was found to be sufficient
for the purpose of our paper.

As shown by multiple authors \cite{Parrish2012, Hu2017, Duchemin2019}, due to the separable form of $\mathbf C$ 
in \cref{eq:fit} the formation of $\xi$ in \cref{eq:xi} can be 
carried out with $\mathcal O(N^3)$ cost.
First,
the formation of $\tilde{\mathbf S}$ scales cubically, 
despite its formal $\mathcal O(N^4)$ scaling,
\begin{equation}
\tilde {S}_{K L}
=
\lambda(\mathbf r_K)
\lambda(\mathbf r_L)
\sum_{\mu}
(\omega_\mu^{K}\omega_\mu^{L})
\sum_\nu{(\omega_\nu^{K}\omega_\nu^{L})}
\label{eq:CC}
\end{equation}
Similarly, the formation of $\mathbf C^\dagger \mathbf Z$ \insertrev{needed for \cref{eq:xi}} can be done as
\begin{equation}
(\mathbf C^\dagger \mathbf Z)_{K,\mathbf r}
=
\sum_{\mu}
(\omega_\mu^{K}\omega_\mu({\mathbf r}))\sum_\nu{(\omega_\nu^{K}\omega_\nu({\mathbf r)})},
\end{equation}
which also scales cubically with system size.

With this decomposition, the two-electron four-center integral tensor can be factorized into \cref{eq:thc}
where we define $\mathbf M$ to be the Coulomb integral between interpolation vectors,
\begin{equation}
M_{KL} = 
\int \mathrm d{\mathbf{r}_1}
\int \mathrm d{\mathbf{r}_2}
\frac{\xi_{K}(\mathbf r_1)\xi_{L}(\mathbf r_2)}{|\mathbf r_1 - \mathbf r_2|}
\label{eq:M}
\end{equation}
As pointed by Lu and Ying \cite{Lu2015}, the evaluation of each Coulomb integral needed to form $\mathbf M$ for translationally invariant systems (e.g., solids) can be done
by fast Fourier transform (FFT) with the $\mathcal O(N\log N)$. In such cases, the formation of $\mathbf M$ scales only cubically with 
the system size. 
Therefore, the overall THC factorization scales cubically with the system size as long as the evaluation of \cref{eq:M} requires only cubic (or less than cubic) work.
In passing, we note that the matrix $\mathbf M$ is  denoted as $\mathbf Z$ in most THC literature. 

However, as suggested in ref. \citenum{Parrish2012}, for molecular systems it is difficult to directly evaluate \cref{eq:M} using quadrature with a finite number of grid points. The challenge is the Coulomb singularity present at $\mathbf r_1 = \mathbf r_2$ which is an integrable singularity. Avoiding this singularity may be possible at the expense of employing very large grids. Usually, the number of grid points required to achieve quantitative accuracy is too large to be practically useful. Instead, one could use FFT to represent $\xi$ with planewaves and perform the Coulomb integrals in the reciprocal space assuming the molecular system of interest is in a large box under periodic boundary conditions. This has been tried in ref \citenum{Dong2018}. However, this would naturally require a pseudopotential treatment of core electrons because it is difficult to represent highly localized core orbitals with planewaves. We must seek  alternatives to perform all-electron calculations.
\subsubsection{Atomic Orbital THC (AO-THC)}
In this work, we will take an alternative route to evaluate $M_{KL}$ which turns out to be identical to ``atomic orbital'' THC (AO-THC) introduced in ref. \citenum{Parrish2012}.
Consider
\begin{equation}
|\xi_{K} ) = \sum_{\mu\nu L}|\mu\nu) C_{\mu\nu}^{L} 
\lambda(\mathbf r_L)
\left(\tilde{\mathbf S}^\dagger\right)_{L K}
\label{eq:xi2}
\end{equation}
It is clear that the projection of this onto $( \mathbf r|$ simply recovers \cref{eq:xi} up to the weight $\lambda(\mathbf r)$ which 
can be neglected if one performs an integration over $\mathbf r$ analytically.
Since we would like to compute $M_{KL}$ based on analytic Coulomb integrals, we employ an RI auxiliary basis set, $|P)$, and express Coulomb integrals in terms of AO and RI integrals. The quantity of interest is 
\begin{equation}
D_{QK} = \sum_P (Q|P)^{-1/2}( P | \xi_{K})=  \sum_{\mu\nu PL} (Q|P)^{-1/2}( P | \mu\nu) C_{\mu\nu}^{L} 
\lambda(\mathbf r_L)
\left(\tilde{\mathbf S}\right)^+_{LK}
\label{eq:Bmat}
\end{equation}
Using this, we have
\begin{equation}
M_{KL} = \sum_Q D_{QK} D_{QL}
\label{eq:M2}
\end{equation}
The bottleneck in the evaluation of $\mathbf M$ is computing an intermediate required for $\mathbf D$, 
\begin{equation}
\sum_{\mu\nu}( P | \mu\nu) C_{\mu\nu}^{L} = \sum_{\mu\nu} ( P | \mu\nu) 
\omega_\mu^{L} \omega_\nu^{L}
\label{eq:bottleneck}
\end{equation}
which scales as $\mathcal O (N_\text{AO}^2 N_\text{X} N_\text{IP})$.
This unfortunately exceeds cubic scaling (without any sparsity consideration). 

We note that this quartic scaling can be reduced to ``asymptotically'' cubic-scaling since the number of shell-pairs, $|\mu\nu)$, scales linearly with system size
for large systems.
Therefore, in this case the scaling for forming $\mathbf D$ is $\mathcal O(N_\text{SP}N_\text{X}N_\text{IP})$.
In addition to the shell-pair screening, a production-level implementation may screen AOs on the interpolation points $\mathbf r_{L}$. The number of significant basis functions at a given point, $\mathbf r_{L}$, should be asymptotically constant as seen in the linear-scaling density functional theory implementations \cite{scuseria1999linear}. With this screening technique, \cref{eq:bottleneck} may be carried out 
with asymptotically quadratic cost ($\mathcal O(cN_\text{IP}N_\text{X})$) where $c$ denotes the number of {\it significant} shell-pairs for a given grid point.
For the purpose of this study, the simpler cubic-scaling algorithm is sufficient and therefore this will be used throughout this paper.

For completeness, we write the \insertnew{matrix $\mathbf M$} in terms of exact two-electron integrals (i.e., RI is not used) as well:
\begin{equation}
M_{KL}
= \sum_{\mu\nu\lambda\sigma}
\sum_{K'L'}
(\mu\nu|\lambda\sigma) C_{\mu\nu}^{K'} C_{\lambda\sigma}^{L'}
\lambda(\mathbf r_{K'})
\lambda(\mathbf r_{L'})
\left(\tilde{\mathbf S}^+\right)_{K'K}
\left(\tilde{\mathbf S}^+\right)_{L'L}
.
\label{eq:thcexact}
\end{equation}
Without considering sparsity, \cref{eq:thcexact}
scales as $\mathcal O(N_\text{AO}^4 N_\text{IP})$ and this is not effective for large-scale applications. 
With sparsity,
the scaling of the construction of $\mathbf M$ can be shown to be cubic.
Namely, the contraction over $\mu$ and $\nu$ to form
\begin{equation}
G_{\lambda\sigma}^{K}
= \sum_{K'}\left(\tilde{\mathbf S}^+\right)_{K'K}(\sum_{\mu\nu} (\mu\nu|\lambda\sigma) C_{\mu\nu}^{K'}\lambda(\mathbf r_{K'}))
\end{equation}
scales as $\mathcal O(N_\text{SP}^2N_\text{IP})$. Similarly, the second step where we contract over $\lambda$ and $\sigma$,
\begin{equation}
M_{KL}
=
\sum_{L'}
\left(\tilde{\mathbf S}^+\right)
_{L'L}
(\sum_{\lambda\sigma}
G_{\lambda\sigma}^{K}
C_{\lambda\sigma}^{L'}
)
,
\end{equation}
scales as
$\mathcal O(N_\text{SP}N_\text{IP}^2)$.
With THC, the generation of the tensor $\mathbf M$ scales worse than the generation of the ERI tensor $(\mu\nu|\lambda\sigma)$,
which asymptotically scales as $\mathcal O(N_\text{SP}^2)$, and the hope is that the prefactor may be smaller with THC
for some system size.

We summarize the discussion of computational cost to form $\mathbf M$ in Table \ref{tab:M}.
\begin{table}[h]
\begin{tabular}{|c|c|c|c|}\hline
Eq. & RI & Shell-pair sparsity     &    Computational cost \\ \hline
\eqref{eq:thcexact}&No & No & $\mathcal O(N_\text{AO}^4 N_\text{IP})$ \\ \hline
\eqref{eq:M2}&Yes & No & $\mathcal O(N_\text{AO}^2 N_\text{X} N_\text{IP})$ \\ \hline
\eqref{eq:thcexact}&No & Yes & $\mathcal O(N_\text{SP}^2 N_\text{IP}) + \mathcal O(N_\text{SP} N_\text{IP}^2)$ \\ \hline
\eqref{eq:M2}&Yes & Yes & $\mathcal O(N_\text{SP} N_\text{X} N_\text{IP}) + \mathcal O(N_\text{X}^3) + \mathcal O(N_\text{X}N_\text{IP}^2) + 
\mathcal O(N_\text{X}^2N_\text{IP})
+\mathcal O({N_\text{IP}}^3)$
\\\hline
\end{tabular}
\caption {Computational cost for forming $\mathbf M$ (\cref{eq:M2},\cref{eq:thcexact}).
Just with \insertnew{shell-pair} sparsity, one can achieve a cubic-scaling algorithm to form $\mathbf M$.
}
\label{tab:M}
\end{table}

\subsubsection{Molecular Orbital THC (MO-THC)}
In some applications of THC, 
one may not need the entire set of MO integrals. For instance,
in the case of RI-MP2, the only required MO block is the occupied-virtual (OV) block .
It is then possible to perform a least-squares fit on the OV block,
which is \revrep{easier to compress}{easier to represent} than the entire MO integrals.
We denote this MO-block-specific THC factorization as MO-THC\cite{Song2017a}.
For later use, 
we define an MO-block-specific quantity $\mathbf D^{[xy]}$ ($x$ and $y$ denote MO blocks which will be o (occupied), v (virtual) or n (AO)),
\begin{equation}
D_{QK}^{[xy]} = \sum_P (Q|P)^{-1/2}( P | \xi_{K}^{[xy]}) 
\label{eq:dmo}
\end{equation}
where
\begin{align}
| \xi_{K}^{[xy]})  &= 
\sum_{p_xq_yL}
| p_xq_y) C_{p_xq_y}^{L} 
\lambda(\mathbf r_{L})
\left(\tilde{\mathbf {S}}^{[xy]}\right)^+
_{LK} \\
& = 
\sum_{p_xq_yL}
\sum_{\mu\nu\lambda\sigma}
| \mu\nu) c_{\mu p_x}c_{\nu q_y}c_{\lambda p_x}c_{\sigma q_y} 
C_{\lambda\sigma}^{L} 
\lambda(\mathbf r_{L})
\left(\tilde{\mathbf {S}}^{[xy]}\right)^+
_{LK}\\
& = 
\sum_{L}
\sum_{\mu\nu\lambda\sigma}
| \mu\nu)
P^{[x]}_{\mu\lambda} P^{[y]}_{\nu\sigma}
\omega_\lambda^{L} \omega_\sigma^{L}
\lambda(\mathbf r_{L})
\left(\tilde{\mathbf {S}}^{[xy]}\right)^+
_{LK}\\
& = 
\sum_{L}
\left(
\sum_{\mu\nu}
| \mu\nu)
\tilde{\omega}^{L}_\mu[x] \tilde{\omega}^{L}_\nu[y]
\right)
\lambda(\mathbf r_{L})
\left(\tilde{\mathbf {S}}^{[xy]}\right)^+
_{LK},
\end{align}
with $C_{\mu p_x}$ being
the MO coefficient for the MO block $x$ and $\left(\tilde{\mathbf {S}}^{[xy]}\right)^+$
is the pseudoinverse of the metric for the pertinent MO block $[xy]$.
The bottleneck in forming $\mathbf D ^{[xy]}$ scales 
as
$\mathcal O(N_\text{SP}N_\text{X}N_\text{IP})$ (i.e., the same as forming $\mathbf D$ as expected).
With this, $\mathbf M$ becomes also MO-block-dependent,
\begin{equation}
M_{KL}^{[wx][yz]} = \sum_Q D_{QK}^{[wx]}D_{QL}^{[yz]}
\end{equation}
and if $[wx]=[yz]$
\begin{equation}
M_{KL}^{[wx]} = \sum_Q D_{QK}^{[wx]}D_{QL}^{[wx]}
\end{equation}
We will specify MO-blocks whenever appropriate. 

\section{Hartree-Fock Theory}
The exact exchange matrix reads
\begin{equation}
K_{\mu\nu} = - \frac12 \sum_{\lambda\sigma}(\mu\lambda|\sigma\nu)P_{\sigma\lambda}
\end{equation}
where the one-particle density matrix, $\mathbf P$, is
\begin{equation}
P_{\sigma\lambda} = 
\sum_i
c_{\sigma i} c_{\lambda i}
\end{equation}
The complexity of evaluating $K_{\mu\nu}$ scales as $\mathcal O(N_\text{AO}^4)$ without any screening.
With proper integral screening, this costs only a quadratic ($\mathcal O(N_\text{SP}^2)$) amount of work.
We will refer to this as the AO-K algorithm.
The RI implementation of the $K$-build asymptotically costs more:
\begin{equation}
K_{\mu\nu}^\text{RI} = -\frac12 
\sum_{iPQ}
(\mu i | P ) (P|Q)^{-1} (Q|i\nu)
\end{equation}
where the final contraction to form $\mathbf K$ scales as $\mathcal O (N_\text{AO}^2 n_\text{occ}N_\text{X})$ (i.e., quartic scaling).

In terms of $\mathbf M$, the evaluation of THC-RI-K is straightforward:
\begin{equation}
K_{\mu\nu}^{\text{THC}}
= -\frac12
\sum_{iKL}
\omega_\mu^{K}
\phi_i^{K}
M_{KL}
\phi_i^{L}
\omega_\nu^{L}
\label{eq:Kisdf}
\end{equation}
where $\phi_i$ is defined in \cref{eq:ao2mo}.
Since every operation in \cref{eq:Kisdf} is a matrix-matrix multiplication \insertnew{(with a carefully chosen contraction ordering)},
the THC-RI-K algorithm scales cubically with system size (i.e., $\mathcal O(N_\text{AO}N_\text{IP}^2)$).
In comparison to RI-K, ISDF improves the asymptotic scaling and 
the hope is that it will show this favorable scaling starting from reasonably small molecules.
When evaluating, \cref{eq:Kisdf} we have two choices for the input $\mathbf M$: MO-THC and AO-THC.
MO-THC uses $\mathbf M^\text{[on]}$ where $\text o$ denotes the occupied MO block and $\text n$ denotes the entire AO block.
AO-THC uses $\mathbf M^{[nn]}$ which does not depend on MOs unlike the first one.

The first one ($\mathbf M^\text{[on]}$) turns out to closely approximate RI-K for a fixed density matrix $\mathbf P$.
From a low-rank perspective,
it is much \revrep{easier to compress}{easier to represent} $(\mu i | i \nu)$ than $(\mu \lambda | \sigma \nu)$. 
This is simply due to the fact that the former has a lower rank to begin with (due to its lower dimension). 
As such, it is expected that the THC factorization works more efficiently with the former integral. 
Indeed, this was found to be the case in the evaluation of the local energy in auxiliary-field quantum Monte Carlo \cite{malone2018overcoming} and solving the Bethe-Salpeter equation \cite{Hu2018}.
However, one must consider additional complications arising from the use of MO-dependent $\mathbf M$ in \cref{eq:Kisdf}.
Since $\mathbf M^\text{[on]}$ is a function of $\mathbf P$, it is necessary to re-derive the Fock matrix properly 
as a derivative of the THC-RI-K energy expression with respect to $\mathbf P$.
This subtlety was not explored in the THC-RI-K implementation paper for planewaves \cite{Hu2017}.
When using orbital gradient driven self-consistent field (SCF) solvers such as geometric direct minimization \cite{Voorhis2002}, ignoring this subtlety leads to
numerically unstable behavior and SCF calculations often do not even converge.
On the other hand, in our experience, the use of other SCF solvers such as Pulay's direct inversion of the iterative subspace (DIIS) \cite{Pulay1980, Pulay1982}
can converge 
provided that the stopping criterion is relatively loose.

Given this, we adopted the AO-THC-RI-K approach which uses MO-independent $\mathbf M = \mathbf M^\text{[nn]}$.
This was found to be generally less accurate than AO-THC for given number of interpolation vectors but it has a few notable strengths:
(1) it is numerically stable and SCF solves can converge tighlty
and (2) the (relatively expensive) formation of $\mathbf M$ needs to occur only once at the beginning of the SCF.

For completeness, we briefly discuss the evaluation of $\mathbf J$.
The exact J-build is
\begin{equation}
J_{\mu\nu} = \sum_{\lambda\sigma}(\mu\nu|\lambda\sigma)P_{\sigma\lambda},
\end{equation}
with the asymptotic scaling of $\mathcal O (N_\text{SP}^2)$.
The RI-J expression reads
\begin{equation}
J_{\mu\nu}^\text{RI} = 
\sum_{\mu\nu\lambda\sigma}
\sum_{PQ}
(\mu \nu | P ) (P|Q)^{-1} (Q|\lambda \sigma) P_{\lambda\sigma},
\end{equation}
which asymptotically scales as $\mathcal O (N_\text{SP} N_\text{X}) + \mathcal O (N_\text{X}^2)$. 
There is also a single cubic scaling step which arises from the formation of $(P|Q)^{-1}$ (i.e., $\mathcal O(N_\text{X}^3)$).
With THC-RI, 
we have
\begin{equation}
J_{\mu\nu}^{\text{THC-RI}} =
\sum_{iKL}
\omega_\mu^{K}
\omega_\nu^{K}
M_{KL}
\phi_i^{L}
\phi_i^{L},
\label{eq:Jisdf}
\end{equation}
which asymptotically scales as 
$\mathcal O (N_\text{SP} N_\text{IP}) +\mathcal O (N_\text{IP}^2) +  \mathcal O (N_\text{IP} n_\text{occ})$.
\insertnew{There is also a cubic scaling step required to form $\mathbf M$.}
\insertnew{As $N_\text{IP}$ is expected to be larger than $N_\text{X}$, the THC-RI-J build is asymptotically more expensive than
the RI-J build.}
\insertnew{We note that just as in THC-RI-K, THC-RI-J can also be done with either AO-THC ($\mathbf M = \mathbf M ^{[nn]}$)
or MO-THC ($\mathbf M = \mathbf M ^{[nn][oo]}$).}

Using RI-J and RI-K at the same time
can result in non-variational HF energies \cite{wirz2017resolution} because the HF energy with both RI-J and RI-K is no longer an upper bound to the exact HF energy.
The origin of this is that $\mathbf J$ is positive-definite whereas $\mathbf K$ is negative-definite.
The RI approximation makes a non-positive error for the energy from $\mathbf J$ but a non-negative error for the energy from $\mathbf K$. 
These two errors do not cancel out and often cause non-positive error overall.
This is the origin of potential non-variationality in RI-HF (and also RI-J itself).
As such,
we investigated THC-RI-K \insertnew{with exact formation of $\mathbf J$} and will present numerical data for it later in this paper.

We present a summary of computational scaling for computing $\mathbf J$ and $\mathbf K$ with different approximations in Table \ref{tab:hf}.
\begin{table}[h]
\begin{tabular}{|c|c|c|c|c|c|}\hline
Algorithm & RI & Sparsity     &   THC & Scaling for $\mathbf J$ & Scaling for $\mathbf K$ \\ \hline
AO & No & No & No &$\mathcal O(M^4)$  & $\mathcal O(M^4)$ \\ \hline
AO & No & Yes & No & $\mathcal O(M^2)$ & $\mathcal O(M^2)$ \\ \hline
RI & Yes & Yes & No & $\mathcal O(M^3)$ & $\mathcal O (M^4)$\\ \hline
THC-RI & Yes & Yes & Yes & $\mathcal O(M^3)$ & $\mathcal O(M^3)$ \\\hline
\end{tabular}
\caption {Computational scaling for forming $\mathbf J$ and $\mathbf K$ in HF.
$M$ here denote a measure for system size that scales linearly with the number of atoms.
\insertnew{Note that RI-J is cubic scaling due to the formation of $(P|Q)^{-1}$ and THC-RI-J is cubic scaling due to the formation
of $\mathbf M$.}
}
\label{tab:hf}
\end{table}

\section{Second-Order M{\o}ller-Plesset Perturbation Theory}
The second-order M{\o}ller-Plesset perturbation theory (MP2) correlation energy with the RI approximation reads 
\begin{equation}
E_\text{MP2}^\text{RI}
=
-\frac14 \sum_{ijab}
\frac{|\langle ij||ab\rangle_\text{RI} |^2}
{
\Delta_{ij}^{ab}
}
\end{equation}
where
$\Delta_{ij}^{ab}$ is the energy denominator,
\begin{equation}
\Delta_{ij}^{ab} = 
\epsilon_a
+\epsilon_b
-\epsilon_i
-\epsilon_j,
\end{equation}
with $\epsilon_p$ being the MO energy of the $p$-th MO
and the RI integrals are
\begin{equation}
\langle ij||ab\rangle_\text{RI}
= \sum_P \left(B_{ia}^P B_{jb}^P- B_{ib}^P B_{ja}^P\right)
\end{equation}
where
\begin{equation}
B_{ia}^P
=
\sum_Q
C_{ia}^Q
(Q|P)^{1/2}
=\sum_Q
(ia|Q)(Q|P)^{-1/2}
\end{equation}
We then define J-like and K-like terms,
\begin{align}
E_\text{MP2-J}^\text{RI}
&=
-\frac12
\sum_{ijab}
\frac{\sum_P(B_{ia}^PB_{jb}^P)\sum_Q(B_{ia}^QB_{jb}^Q)}
{\Delta_{ij}^{ab}} 
\\
E_\text{MP2-K}^\text{RI}
&=
\frac12
\sum_{ijab}
\frac{\sum_P(B_{ia}^PB_{jb}^P)\sum_Q(B_{ib}^QB_{ja}^Q)}
{\Delta_{ij}^{ab}} 
\end{align}
and evidently $E_\text{MP2} = E_\text{MP2-J} + E_\text{MP2-K}$.
The computational scaling of both RI-MP2-J and RI-MP2-K correlation energy evaluation is $\mathcal O (n_\text{occ}^2n_\text{vir}^2N_\text{X})$.

With the Laplace transformation (LT) applied to the energy denominator \cite{haser1992laplace}, 
the correlation energy can be rewritten as
\begin{equation}
E_\text{MP2}^\text{RI}
=
-\frac14\int_0^\infty \mathrm d t\:
\sum_{ijab}
{|\langle ij||ab\rangle_\text{RI} |^2}
 e^{-t \Delta_{ij}^{ab}}
\end{equation}
where
in addition to 
the sum over MO indices
there is also a one-dimensional (1D) quadrature over $t$.
The number of quadrature points does not scale with system size so this will be left out in 
the cost analysis below \cite{haser1992laplace}.
With this trick, an interesting cost reduction occurs in the evaluation of the RI-MP2-J correlation energy, (with explicit spin-sum)
\begin{equation}
E_\text{MP2-J} ^\text{RI} =
-\frac12
\int_0^\infty \mathrm d t\:
\sum_{\sigma_1\sigma_2}
\sum_{PQ}X_{PQ}^{\sigma_1}(t) X_{PQ}^{\sigma_2}(t)
\label{eq:mp2-j}
\end{equation}
where we performed the sum over MO indices first, and
\begin{equation}
X_{PQ}^\sigma(t)
= 
\sum_{i_\sigma a_\sigma}
B_{i_\sigma a_\sigma}^P B_{i_\sigma a_\sigma}^Q
 e^{-t (\epsilon_{a_\sigma} - \epsilon_{i_\sigma} )}.
\end{equation}
The formation of $\mathbf X^\sigma$ is the bottleneck in evaluating the RI-MP2-J correlation energy,
which scales as
$\mathcal O(n_\text{occ}n_\text{vir}N_\text{X}^2)$. This quartic scaling algorithm
for the evaluation of RI-MP2-J correlation energy has been used
in the scaled opposite-spin (SOS)-MP2 method \cite{jung2004scaled}. The opposite-spin correlation energy involves only the RI-MP2-J term and therefore can be evaluated with this algorithm at a reduced scaling. 

The RI-MP2-K correlation energy, which appears only in the same-spin correlation energy, reads (with explicit spin-sum)
\begin{equation}
E_\text{MP2-K} ^\text{RI}=
\frac12
\int_0^\infty \mathrm d t\:
\left[
\sum_{\sigma}
\sum_{i_{\sigma} a_{\sigma}}\sum_{j_{\sigma} b_{\sigma}}
\sum_P(B_{i_\sigma a_\sigma}^PB_{j_\sigma b_\sigma}^P)\sum_Q(B_{i_\sigma b_\sigma}^QB_{j_\sigma a_\sigma}^Q)
 e^{-t \Delta_{i_\sigma j_\sigma}^{a_\sigma b_\sigma}}
 \right]
\end{equation}
Summing over the MO indices first results in a worse computational scaling compared to the usual RI-MP2-K algorithm without the Laplace trick. Therefore, there is not much computational benefit for performing LT-RI-MP2.
One may work directly with AO quantities and exploit sparsity of integrals in the AO basis set to lower the asymptotic scaling as is done in LT-AO-MP2 \cite{ayala1999linear} but 
this involves large prefactors and 
we do not discuss this here.

With THC-RI, further cost reductions can be achieved for both RI-MP2-J and RI-MP2-K.
The THC-RI-MP2-J correlation energy reads
\begin{equation}
E_\text{MP2-J}^\text{THC-RI}
=
-\frac12
\int_0^\infty \mathrm d t\:
\sum_{KLMN}
M_{KL}^{[ov]}
M_{MN}^{[ov]}
W_{KM}(t)
\tilde{W}_{KM}(t)
W_{LN}(t)
\tilde{W}_{LN}(t)
\label{eq:isdf-mp2-j}
\end{equation}
where
\begin{equation}
W_{KM} (t) = 
\sum_i 
\phi_i^{K}
\left(\frac t2\right) 
\phi_i^{M} 
\left(\frac t2\right) 
=
\sum_i 
\phi_i^{K}(0)
\phi_i^{M} 
\left(t\right) 
\label{eq:woo}
\end{equation}
and
\begin{equation}
\tilde{W}_{KM} (t) = 
\sum_a \phi_a^{K}
\left(\frac t2\right) 
\phi_a^{M}\left(\frac t2\right) 
=
\sum_a \phi_a^{K}
(0)
\phi_a^{M}\left(t\right) 
\label{eq:wvv}
\end{equation}
The cost of evaluating \cref{eq:isdf-mp2-j} is $\mathcal O(N_\text{X}N_\text{IP}^2)$ (i.e., cubic-scaling with system size).
This cubic-scaling is due to the fact that \cref{eq:isdf-mp2-j} involves only matrix multiplications
and the formation of $\mathbf W$ and $\tilde{\mathbf W}$ scales also cubically with system size.
Therefore, with THC, the evaluation of RI-MP2-J can be completed with a cubic cost \cite{Song2016} in contrast to its original quartic scaling in \cref{eq:mp2-j}. 

The THC-RI-MP2-K correlation energy is given as
\begin{align}\nonumber
E_\text{MP2-K}^\text{THC-RI}
=&
-\frac12
\int_0^\infty \mathrm d t\:
\sum_{KLMN}
\sum_{iab}
\bigg(
M_{KL}^\text{[ov]}
M_{MN}^\text{[ov]}
\phi_i^{K}\left(\frac{t}{2}\right)
\phi_i^{M}\left(\frac{t}{2}\right)
\phi_a^{K}\left(\frac{t}{2}\right)
\phi_a^{N}\left(\frac{t}{2}\right)\\
&\phi_b^{L} \left(\frac{t}{2}\right)
\phi_b^{M} \left(\frac{t}{2}\right)
W_{LN}(t)\bigg)
\label{eq:isdfmp2k2}
.
\end{align}
Assuming that $N_\text{IP} > N_\text{X} > n_\text{vir} > n_\text{occ}$ and no sparsity is considered, a different contraction strategy (from that of MP2-J) leads to a potentially more efficient (in terms of both storage and operations) algorithm. 
Starting from \cref{eq:isdfmp2k2}, we introduce another intermediate, 
\begin{equation}
\tilde{B}_{ib}^{\hat {Q}}(t)
= \sum_{K} M_{K L}^\text{[ov]} \phi_i^{K}(t)\phi_b^{K}(t)
\end{equation}
where $\mathbf M$ is given in \cref{eq:M2}. The formation of this intermediate scales as $\mathcal O(n_\text{occ}n_\text{vir}N_\text{IP}^2)$ and the storage requirement scales as  $\mathcal O(n_\text{occ}n_\text{vir}N_\text{IP})$.
Furthermore, we define a second intermediate that can be obtained from $\tilde{\mathbf B}$.
It is defined as
\begin{equation}
\tilde{C}_{iN}^{L}(t, s)
= \sum_b \tilde{B}_{ib}^{L}(t) \phi_b^{N}(s), 
\label{eq:thcC}
\end{equation}
which requires $\mathcal O(n_\text{occ}n_\text{vir}N_\text{IP}^2)$ operations 
and $\mathcal O(n_\text{occ}N_\text{IP}^2)$ storage (potentially $\mathcal O(N_\text{IP}^2)$ with a direct algorithm).
With this, we rewrite the THC-RI-MP2-K correlation energy:
\begin{equation}
E_\text{MP2-K}^\text{THC-RI}
=
-\frac12
\int_0^\infty \mathrm d t\:
\sum_{iNL}
\tilde{C}_{iN}^{L}\left(\frac t2,\frac t2\right)
\tilde{C}_{iL}^{N}\left(\frac t2,\frac t2\right)
W_{LN}(t)
\end{equation}
which scales cubically with system size.
For the numerical simulations presented below, we adopted this algorithm.
We note that summing over all the MO indices in MP2-K leads to a triple sum over interpolation points with $\mathcal O(N_\text{IP}^4)$ scaling. This alternative algorithm could be more efficient with sparsity.
It will be interesting to compare these two algorithms with careful sparsity consideration in the future.
We note that this algorithm is more or less the same as the THC-RI-MP2 algorithm with restricted orbitals proposed by Mart{\'i}nez and co-workers \cite{Hohenstein2012, KokkilaSchumacher2015}.
In Algorithm \ref{algo:mp2}, we illustrate step-by-step our algorithm for evaluating the THC-RI-MP2
same-spin correlation energies.
The resulting algorithm scales as $\mathcal O(N_\text{IP}^3)$ and $\mathcal O(N_\text{IP}^3n_\text{occ})$ for THC-RI-MP2-J and THC-RI-MP2-K, respectively. Furthermore, the storage requirement is only quadratic with system size for both energy contributions.

\begin{algorithm}[H]
\label{algo:mp2}
\SetAlgoLined
\For{$t=1$ \KwTo $n_t$\tcp*{Loop over Laplace points.}}
{
Form $\omega_i^{K}(t/2) = \omega_i^{K}e^{t/2\epsilon_i}$\tcp*{$\mathcal O(n_\text{occ}N_\text{IP})$}
Form $\omega_a^{K}(t/2) = \omega_a^{K}e^{-t/2\epsilon_a}$\tcp*{$\mathcal O(n_\text{vir}N_\text{IP})$}
Form $W_{K L} (t)$ (\cref{eq:woo})\tcp*{$\mathcal O(n_\text{occ}N_\text{IP}^2)$}
Form $\tilde{W}_{K L} (t)$ (\cref{eq:wvv})\tcp*{$\mathcal O(n_\text{vir}N_\text{IP}^2)$}
Compute $A_{KL} = \sum_{M} W_{KM} \tilde{W}_{KM} M_{ML}$\tcp*{$\mathcal O(N_\text{IP}^3)$}
Accumulate $E_J$ --= $0.5 * w_t * \sum_{KL} A_{KL}A_{LK}$\tcp*{$\mathcal O(N_\text{IP}^2)$}
\For{$i=1$ \KwTo $n_\text{occ}$\tcp*{(Parallel) loop over occupied orbitals.}}{
Compute $\tilde{C}_{K L} = \sum_{M} \tilde{W}_{K M} \omega_i^{M}(t/2) M_{\hat {R}L}$\tcp*{$\mathcal O(N_\text{IP}^3n_\text{occ})$}
Accumulate $E_K$ += $0.5 * w_t * \sum_{KL} \tilde{C}_{K L} \tilde{C}_{L K} W_{KL}$\tcp*{$\mathcal O(N_\text{IP}^2n_\text{occ})$}
}
}
\caption{THC-RI-MP2 algorithm.
$w_t$ denotes a Laplace quadrature weight and $E_J$ and $E_K$ denote
the THC-RI-MP2-J and THC-RI-MP2-K same-spin correlation energies, respectively.
}
\end{algorithm}

In passing we note that we used $\mathbf M^{[ov]}$ as opposed to $\mathbf M^{[nn]}$.
$\mathbf M^{[ov]}$ is, due to its lower rank, \revrep{easier to compress}{easier to represent} and fit. As such, the accuracy of MO-THC is
usually higher than that of AO-THC\cite{Song2017a} for a fixed number of interpolation points compared to when compressing $\mathbf M^{[nn]}$. \insertrev{This higher accuracy comes with more difficult derivative expression and implementation of response theory.}

Lastly, we briefly discuss the use of THC-RI for recently developed regularized MP2 ($\kappa$-MP2). 
Along with orbital optimization (OOMP2),
$\kappa$-OOMP2 has been shown to be useful
for thermochemistry, non-covalent interaction, and barrier heights \cite{lee2018regularized}.
More recently, it was applied to distinguishing artificial and essential symmetry breaking \cite{lee2019distinguishing} and singlet biradicaloids \cite{lee2019two}.
Furthermore, using $\kappa$-OOMP2 orbitals was shown to improve the performance of MP3 greatly \cite{bertels2019third}.
As such, $\kappa$-OOMP2 (and $\kappa$-MP2) is a useful electron correlation model and it is worthwhile to see if the same scaling reduction can be applied to it.
The RI-$\kappa$-MP2 correlation energy is
\begin{align}
E_\text{$\kappa$-MP2}^\text{RI}
&=
-\frac14
\sum_{ijab}
\frac{|\langle ij||ab\rangle_\text{RI} |^2}{\Delta_{ij}^{ab}}
(1-e^{-\kappa\Delta_{ij}^{ab}})^2
\\
&=
-\frac14\int_0^\infty \mathrm d t\:
\sum_{ijab}
{|\langle ij||ab\rangle_\text{RI} |^2}
 e^{-t \Delta_{ij}^{ab}} 
(1-e^{-\kappa\Delta_{ij}^{ab}})^2 \\
 &=
 -\frac14\int_0^\infty \mathrm d t\:
\sum_{ijab}
{|\langle ij||ab\rangle_\text{RI} |^2}
 (e^{-t \Delta_{ij}^{ab}} -2e^{-(t+\kappa)\Delta_{ij}^{ab}}+e^{-(t+2\kappa)\Delta_{ij}^{ab}}) 
\end{align}
The last expression merely suggests that
the RI-$\kappa$-MP2 correlation energy may be computed as
three LT-RI-MP2 energy evaluations with shifted $t$ values and a single MO-integral evaluation. 
The application of ISDF to this expression is essentially identical to what was discussed above for THC-RI-MP2.
Therefore, the scalings of THC-RI-$\kappa$-MP2-J and THC-RI-$\kappa$-MP2-K are cubic and quartic, repsectively.

A summary of computational scaling and storage requirement for evaluating MP2-J and MP2-K correlation energies
is given in Table \ref{tab:mp2}.
\begin{table}[h]
\begin{tabular}{|c|c|c|c|c|}\hline
RI & LT & THC & Scaling for MP2-J & Scaling for MP2-K \\ \hline
No & No & No &$\mathcal O(M^5)$  & $\mathcal O(M^5)$\\ \hline
Yes & Yes & No & $\mathcal O(M^4)$  & $\mathcal O(M^5)$\\ \hline
Yes & Yes & Yes & $\mathcal O(M^3)$ & $\mathcal O (M^4)$\\\hline
\end{tabular}
\caption {Computational scaling for evaluating MP2-J and MP2-K correlation energies.
$M$ here denote a measure for system size that scales linearly with the number of atoms.
}
\label{tab:mp2}
\end{table}

\section{Third-Order M{\o}ller-Plesset Perturbation Theory}\label{sec:mp3}
As noted in ref. \citenum{Hohenstein2012}, a more drastic scaling reduction is possible for the third-order M{\o}ller-Plesset perturbation theory (MP3) correlation energy evaluation.
The MP3 correlation energy reads
\begin{equation}
E_\text{MP3} = 
E_\text{vv}^{(3)}
+E_\text{oo}^{(3)}
+E_\text{ov}^{(3)}
\label{eq:mp3}
\end{equation}
where the spin-orbital expressions for each term are:
\begin{align}\label{eq:mp3vv}
E_\text{vv}^{(3)} &= \frac 18
\sum_{ijabcd} t_{ij}^{ab} \langle ab || cd \rangle t_{ij}^{cd}\\ \label{eq:mp3oo}
E_\text{oo}^{(3)} &=
\frac 18
\sum_{ijklab} t_{ij}^{ab} \langle ij || kl \rangle t_{kl}^{ab}\\ \label{eq:mp3ov}
E_\text{ov}^{(3)} &= -
\sum_{ijabkc} t_{ij}^{ab} \langle ic || kb \rangle t_{kj}^{ac}
\end{align}
which scales as $\mathcal O(n_\text{occ}^2n_\text{vir}^4)$, $\mathcal O(n_\text{occ}^4n_\text{vir}^2)$, and $\mathcal O(n_\text{occ}^3n_\text{vir}^3)$ with system size (i.e., sextic scaling), respectively.
The application of RI does not reduce the asymptotic computational cost but may help to reduce the storage requirement.

From the THC point of view, the MP3 correlation energy poses an additional challenge that was not present in HF and MP2.
Namely, the RI-MP3 correlation energy requires the representation of oo, vv, and ov blocks of the RI MO-integrals.
Based on rank considerations, it is expected that the oo-block is the easiest to compress and the vv-block is the most difficult to compress.
In passing, we note that $E_\text{vv}^{(3)}$, $E_\text{oo}^{(3)}$, and $E_\text{ov}^{(3)}$
have a total of two, two, and four, respectively, unique contributions depending on the spin-blocks of the first $\mathbf t$, the middle integral, and the second $\mathbf t$.
Such unique contributions are summarized in Table \ref{tab:mp3so}.
\begin{table}[h]
\begin{tabular}{|c|c|c|c|}\hline
Contribution & First $\mathbf t$ & Middle integral & Second $\mathbf t$\\ \hline
$E_\text{vv-SS}^{(3)}$ & SS & SS & SS\\ \hline
$E_\text{vv-OS}^{(3)}$ & OS & OS & OS\\ \hline
$E_\text{oo-SS}^{(3)}$ & SS & SS & SS\\ \hline
$E_\text{oo-OS}^{(3)}$ & OS & OS & OS\\ \hline
$E_\text{ov-SS-SS-SS}^{(3)}$ & SS & SS & SS\\ \hline
$E_\text{ov-OS-OS-OS}^{(3)}$ & OS & OS & OS\\ \hline
$E_\text{ov-OS-OS-SS}^{(3)}$ & OS & OS & SS\\ \hline
$E_\text{ov-OS-SS-OS}^{(3)}$ & OS & SS & OS\\ \hline
\end{tabular}
\caption {
Individual energy contributions in $E_\text{MP3}$ categorized by the spin-blocks of the first $\mathbf t$, the middle integral, and
the second $\mathbf t$. SS means ``same-spin'' while OS means ``opposite-spin''.
}
\label{tab:mp3so}
\end{table}

Applying LT twice to \cref{eq:mp3vv}, \cref{eq:mp3oo}, and \cref{eq:mp3ov} and using the RI approximation and following the same THC factorization described in ref. \citenum{Hohenstein2012}, it is straightforward to show that
every energy contribution can be computed with a quartic cost and quadratic storage.
As an example, 
we discuss the application of THC-RI to $E_\text{vv}^{(3)}$.
As described in Table \ref{tab:mp3so}, we note that 
$E_\text{vv}^{(3)}$ and $E_\text{oo}^{(3)}$ have only two distinct spin-block contributions: same-spin (SS) and opposite-spin (OS).
Namely, 
\begin{equation}
E_\text{vv}^{(3)} = E_\text{vv-SS}^{(3)} + E_\text{vv-OS}^{(3)}
\end{equation}
where
\begin{align}
E_\text{vv-SS}^{(3)} &= \frac 18\sum_{\sigma\in\{\alpha, \beta\}} 
\sum_{\substack{i_\sigma j_\sigma \\a_\sigma b_\sigma c_\sigma d_\sigma } }
t_{i_\sigma j_\sigma }^{a_\sigma b_\sigma } \langle a_\sigma b_\sigma  || c_\sigma d_\sigma  \rangle t_{i_\sigma j_\sigma }^{c_\sigma d_\sigma }\\
E_\text{vv-OS}^{(3)} &= \frac14
\sum_{\substack{\sigma,\sigma'\in\{\alpha,\beta\}\\\sigma\ne\sigma'} }
\sum_{\substack{i_\sigma j_{\sigma'} \\a_{\sigma} b_{\sigma'} c_\sigma d_{\sigma'} }}
t_{i_\sigma j_{\sigma'} }^{a_\sigma b_{\sigma'} } 
\langle a_\sigma b_{\sigma'}  | c_\sigma d_{\sigma'}
\rangle t_{i_\sigma j_{\sigma'} }^{c_\sigma d_{\sigma'} }
\end{align}
We take an approach where we (1) sum over all the MO indices first, (2) loop over an interpolation point index, (3) store everything for a given index, and (4) accumulate the energy contributions. 
We loop over $\hat U$ in \cref{eq:vvSS} and compute intermediates accordingly:
\begin{equation}
E_\text{vv-SS}^{(3)}
=
\int_t\int_s
\sum_{K L M}
\left(
A_{K L}^{M}(t) B_{K L}^{M}(t,s)
- 
(A1)^{M}_{K L}(t,s)
(A2)^{M}_{K L}(t,s)
(A3)^{M}_{K L}(t,s)
\right)
\label{eq:vvSS}
\end{equation}
where
\begin{align}
A1_{KL}^M(t,s) &= \sum_{M} W_{K M}(t+s) M^\text{[ov]}_{M \hat U} \tilde{W}_{M L}(s)\\
A2_{KL}^M(t,s) &= \sum_{M} M^\text{[ov]}_{K M} W_{M \hat U}(t+s) \tilde{W}_{M L}(t)\\
A3_{KL}^M(t,s) &= \sum_{M} \tilde{W}_{K M}(t) \tilde{W}_{M \hat U}(s) M^\text{[vv]}_{M L}
\end{align}
This strategy ensures the quartic computational cost and quadratic storage cost without considering sparsity.
More or less identical strategies can be applied to $E_\text{oo-SS}^{(3)}$ and $E_\text{oo-OS}^{(3)}$
and the resulting expressions exhibit quartic scaling.
We will not explicitly write down $E_\text{oo-SS}^{(3)}$ and $E_\text{oo-OS}^{(3)}$ because
they may be obtained by replacing virtuals with occupieds and vice versa in $E_\text{vv-SS}^{(3)}$ and $E_\text{vv-OS}^{(3)}$.
Likewise, the ov contribution can be obtained with THC-RI using similar strategies described here.

This strategy, however, comes with a high prefactor due to the double LT (i.e., a prefactor of $n_t^2$).
With 7 Laplace quadrature points \cite{jung2004scaled}, this prefactor becomes 49. 
As such, we found it to be difficult to observe computational benefits for medium-sized molecules this way.
The large prefactor from double LT can be {\it completely} removed if one starts from THC-factorized first-order MP amplitudes:
\begin{equation}
t_{ij}^{ab} = \sum_{K L} 
\left(
\tau^{K}_i
\tau^{K}_a
T_{K L}
\tau^{L}_j
\tau^{L}_b
-
\tau^{K}_i
\tau^{K}_b
T_{K L}
\tau^{L}_j
\tau^{L}_a
\right)
\label{eq:tthc}
\end{equation}
where $\tau^{K}_p$ is some basis functions evaluated on a 3-space point $\mathbf r_{K}$
and $T_{K L}$ is defined to make  \cref{eq:tthc} true in the least-squares sense.
\insertrev{Factorizing amplitudes has been explored by several groups in the context of coupled-cluster methods \cite{Hohenstein2012a,benedikt2013tensor,Shenvi2014,schutski2017tensor}. Our approach is essentially the same as
that of ref. \citenum{Hohenstein2012a}, and we apply it to the evaluation of MP3 energy.}
The choice for the $\tau$ basis functions may not seem immediately obvious and
for simplicity we used the MOs $\phi_p^{K}$ and the same set of interpolation points as for the integral factorization.
This turns out to work reasonably well \insertrev{for small basis sets} although it takes a few more interpolation points than the double LT algorithm described above for a fixed target accuracy.

The factorization in \cref{eq:tthc} seems more reasonable when considering
\begin{equation}
t_{ij}^{ab} = 
\int_t
\left(
\phi^{K}_i (t)
\phi^{K}_a (t)
\lambda_K
M_{K L}^{[ov]}
\lambda_L
\phi^{L}_j (t)
\phi^{L}_b (t)
-
\phi^{K}_i (t)
\phi^{K}_b (t)
\lambda_K
M_{K L}^{[ov]}
\lambda_L
\phi^{L}_j (t)
\phi^{L}_a (t)
\right).
\label{eq:tthc2}
\end{equation}
The structure of $t_{ij}^{ab}$ is already factorized in \cref{eq:tthc2} with a summation over the quadrature points.
Our goal is then to find $\mathbf T$ which best satisfies
\begin{equation}
\sum_{K L}
\phi^{K}_i
\phi^{K}_a
\lambda_K
T_{K L}
\lambda_L
\phi^{L}_j 
\phi^{L}_b
=
\int_t
\sum_{K L}
\left(
\phi^{K}_i (t)
\phi^{K}_a (t)
\lambda_K
M_{K L}^{[ov]}
\lambda_L
\phi^{L}_j (t)
\phi^{L}_b (t)
\right)
\end{equation}
We then define a time-dependent version of \cref{eq:fit},
\begin{equation}
C^{K}_{ia}(t) = \phi^{K}_{i} (t) \phi^{K}_{a}(t)
\end{equation}
which is an $n_\text{occ}n_\text{vir}$-by-$N_\text{IP}$ matrix.
Then,
the least-squares solution for $\mathbf T$ is
\begin{equation}
\mathbf T = 
(\tilde{\mathbf S}^\text{[ov]}(0))^{+}
\left(
\int_t
\tilde{\mathbf S}^\text{[ov]}\left(\frac t2\right)
\mathbf M^\text{[ov]}
\tilde{\mathbf S}^\text{[ov]}\left(\frac t2\right)
\right)
(\tilde{\mathbf S}^\text{[ov]}(0))^{+}
\label{eq:tfacfinal}
\end{equation}
where
the time-dependent metric is given as (analogously to \cref{eq:metric}),
\begin{equation}
\tilde{S}^\text{[ov]}_{KL}(t) = \sum_{ia} \lambda_K C^{K}_{ia}(t) C^{L}_{ia}(t)\lambda_L
\end{equation}
Similar to \cref{eq:CC},
the evaluation of $\tilde{\mathbf S}^\text{[ov]}(t)$
can be done at a cubic cost. 
Furthermore, it is also generally a ill-conditioned, singular matrix so a threshold of $10^{-12}$ 
was used in the drop tolerance for calculating its pseudoinverse.
Therefore, obtaining the THC-factorized first-order MP amplitudes
scales only cubically with system size.
With $\mathbf T$,
one can remove the double quadrature loop in Algorithm \ref{algo:mp3-vv}
and thereby the large prefactor of $n_t^2$ no longer exists.
Our numerical results are obtained with this additional factorization of $\mathbf t$.

The detailed algorithm for the MP3-VV correlation energy in our final THC-RI-MP3 formulation is presented in Algorithm \ref{algo:mp3-vv}.

\begin{algorithm}[H]
\label{algo:mp3-vv}
\SetAlgoLined
Call Subroutine FormT \tcp*{Form the core tensor $\mathbf T$}
\For{$\hat{U}=1$ \KwTo $N_\text{IP}$\tcp*{Loop over interpolation points.}}
{
Form $A_{K L} = \sum_{M} \tilde{W}_{K M}M^\text{[vv]}_{M \hat U} \tilde{W}_{M L}$\tcp*{$\mathcal O(N_\text{IP}^4)$} 
Form $B_{K L} = \sum_{M N} T_{K M}
\tilde{W}_{M \hat U}
\tilde{W}_{N \hat U}
W_{M N}
T_{N L}$\tcp*{$\mathcal O(N_\text{IP}^4)$}
E += $\sum_{KL} A_{K L} B_{K L}$\tcp*{$\mathcal O(N_\text{IP}^3)$}
Form $A1_{KL} = \sum_{M} W_{K M} T_{M \hat U} \tilde{W}_{M L}$\tcp*{$\mathcal O(N_\text{IP}^4)$}
Form $A2_{KL} = \sum_{M} T_{K M} W_{M \hat U}\tilde{W}_{M L}$\tcp*{$\mathcal O(N_\text{IP}^4)$}
Form $A3_{KL} = \sum_{M} \tilde{W}_{K M} \tilde{W}_{M \hat U} M^\text{[vv]}_{M L}$\tcp*{$\mathcal O(N_\text{IP}^4)$}
$E$ --= $\sum_{KL} A1_{K L}A2_{K L}A3_{K L}$\tcp*{$\mathcal O(N_\text{IP}^3)$}
}
\caption{THC-RI-MP3 algorithm for evaluating $E_\text{vv-SS}^{(3)}$.
}
\end{algorithm}

We present a summary of computational scaling and storage for MP3 in Table \ref{tab:mp3}.
\begin{table}[h]
\begin{tabular}{|c|c|c|c|}\hline
RI & LT & THC & Scaling for MP3\\ \hline
No & No & No &$\mathcal O(M^6)$\\ \hline
Yes & Yes & No & $\mathcal O(M^6)$ \\ \hline
Yes & Yes & Yes & $\mathcal O(M^4)$\\ \hline
\end{tabular}
\caption {Computational scaling and storage cost for evaluating MP3 correlation energies.
$M$ here denotes a measure of molecule size that scales linearly with the number of atoms.
}
\label{tab:mp3}
\end{table}

\section {Computational Details}
We consider two prototypical quantum chemistry benchmark sets in this work:
(1) the W4-11 set for total atomization energies \cite{Karton2011} and (2) the A24 set for non-covalent interaction energies \cite{Rezac2013}.
W4-11 contains a total of 140 data points and spans an energy range from 2.67 kcal/mol to 1007.9 kcal/mol.
A24 contains 24 data points which accounts for energies ranging from 0.37 kcal/mol to 6.53 kcal/mol.
It is crucial to examine the utility of THC on both of the energy scales. 
MP2 and MP3 calculations on the A24 set were all counterpoise corrected \cite{boys1970calculation}.

We examine the effect of the basis set on the accuracy of THC-RI calculations
for HF,  $\omega$B97X-V \cite{mardirossian2014omegab97x} (a range-separated hybrid functional), MP2, and MP3.
This was done by considering 
commonly used Dunning's cc-pVDZ, cc-pVTZ, aug-cc-pVDZ, and aug-cc-pVTZ basis sets \cite{Dunning1989,woon1993gaussian, kendall1992electron}.
The pertinent RI basis sets for each AO basis set were used \cite{weigend2002efficient}.
\insertnew{Although there are different RI basis sets available for RI-K, we used RI basis set made for RI-MP2 for all methods.
This was sufficient for the purpose of this study.}
We note that there has 
been no report of THC calculations using augmented bases previously.

In every calculation, we allowed for spin-unrestriction.
For HF and $\omega$B97X-V calculations, we performed  wavefunction internal stability analysis\cite{Seeger1977} to ensure 
the local stability of obtained solutions.
For RI-K and THC-RI-K simulations combined with HF and $\omega$B97X-V, 
we use converged exact SCF solutions as a guess
and did not perform the stability analysis for those.
We used 99 radial points and 590 Lebedev angular points for all $\omega$B97X-V calculations.
Geometric direct minimization (GDM) \cite{Voorhis2002}
was used for the SCF optimizer
and ${10}^{-5}$ to ${10}^{-6}$ for the root-mean-square of 
the orbital gradient
was used for the convergence criterion. This is sufficient for single-point calculations.
All correlated wavefunction calculations are based on exact SCF solutions and correlate all electrons.

The CVT procedure was performed with the parent grid of SG-1\cite{gill1993standard}, which contains
50 radial points and 194 Lebedev angular points for each atom. This was found to be sufficient for 
the systems considered in this paper.
We used the standard 7-point Laplace quadrature grid as suggested in ref. \citenum{jung2004scaled}.

All calculations were performed with a development version of
Q-Chem \cite{Shao2015}
which involves the use of
(1) an integral library (\texttt{libqints}) which computes AO integrals for gaussian-type-orbitals,
(2) an integral transformation library (\texttt{libposthf}) which produces RI-MO integrals,
(3) a general quadrature library (\texttt{libgrid}) which takes care of the Becke grid generation,
(4) an SCF library (\texttt{libgscf}) which performs an SCF calculation for arbitrary types of orbitals,
(5) a generalized many-body perturbation theory library (\texttt{libgmbpt}) which performs MP2 and MP3,
and
(6) an ISDF/THC library (\texttt{libisdf}) which implements THC-RI algorithms for various methods.
Except \texttt{libqints}, those libraries were developed over a number of publications
including the present work
and interested readers are referred to refs. \citenum{lee2017coupled, lee2018open, lee2018regularized, lee2019distinguishing, lee2019two, lee2018generalized, lee2019kohn, lee2019auxiliary} for further details.

\section{Results and Discussion}
For various basis sets,
we will investigate the utility of the THC factorization in conjunction 
with the ISDF point selection.
We assess the accuracy of THC-RI methods by studying the root-mean-square-deviation (RMSD) and maximum unsigned deviation (MAX) from the exact RI results for
both W4-11 and A24 sets. RMSD shows whether the underlying distribution of relative energies is changed by THC and MAX highlights a chemical example that is least accurately described by THC. 
We note that for MP2 and MP3 the exact results employ both RI and LT. The error from LT was found to be negligible in the systems studied in this paper.

We use $c_\text{ISDF}$ to denote the number of interpolation points (i.e., $N_\text{IP}$) such that
they are a multiple of the number of auxiliary basis functions, $N_\text{X}$:
\begin{equation}
N_\text{IP} = c_\text{ISDF} \times N_\text{X}
\end{equation}
We scanned over $c_\text{ISDF} \in [2.0, 10.0]$ in the case of THC-RI-K and
$c_\text{ISDF} \in [2.0, 5.0]$ in the case of MP2 and MP3 with an increment of 0.5.

\subsection{THC-RI-K for HF and $\omega$B97X-V}
As mentioned before, the evaluation of exact exchange
with MO-THC
requires modification to the Fock matrix evaluation.
Because of this, we chose to use AO-THC for exact exchange although a more compact THC factorization should be possible with MO-THC.
As we will see from numerical results, AO-THC still can be converged to numerically exact RI-K energies
with a sufficiently large number of interpolation points.
\begin{figure}[h!]
\includegraphics[scale=0.85]{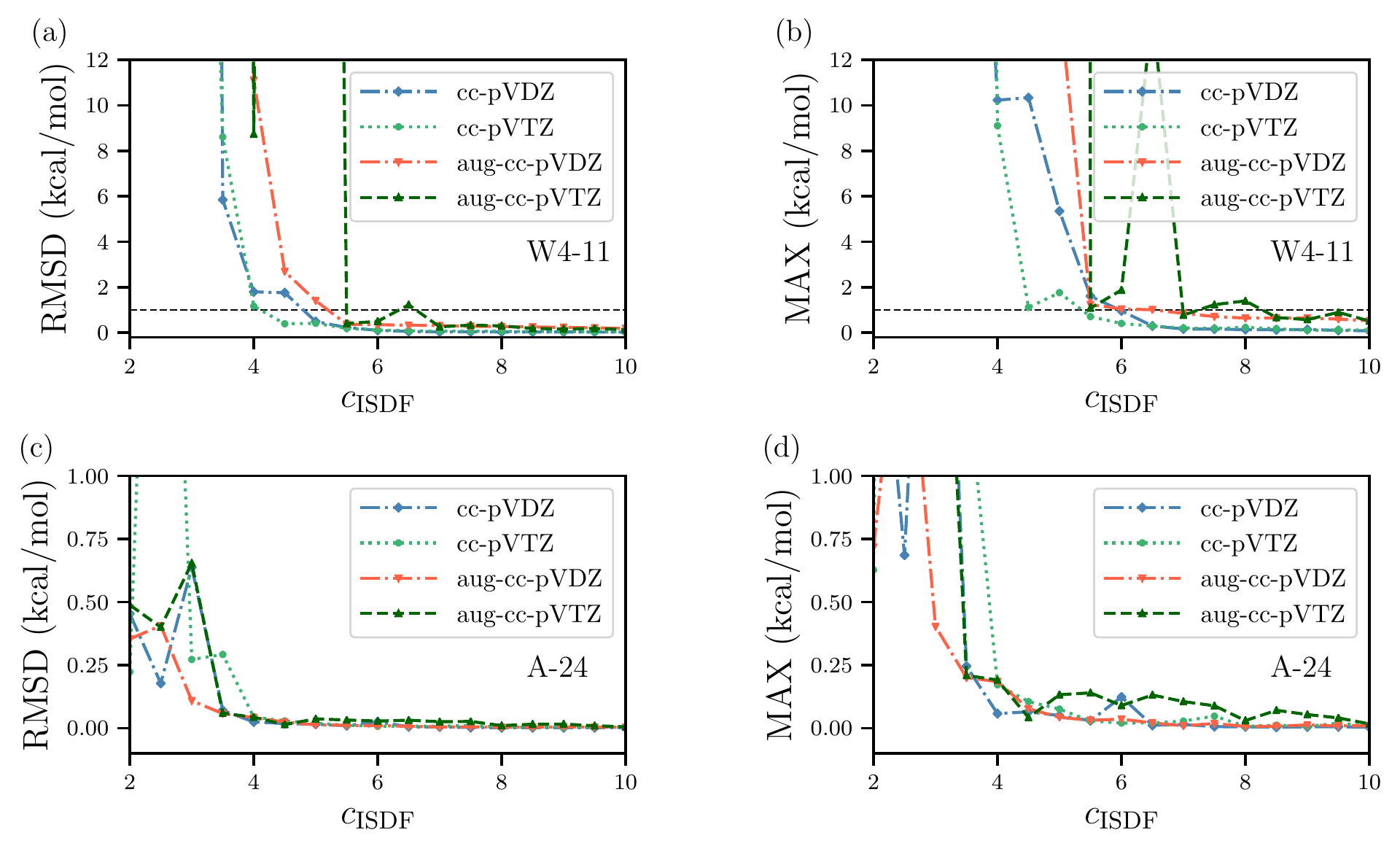}
\caption{\label{fig:hf}
THC-RI-K HF errors (kcal/mol) with respect to
exact RI-K HF for various basis sets:
(a) root-mean-square-deviation (RMSD) for W4-11
(b) maximum unsigned deviation (MAX) for W4-11,
(c) RMSD for A24, and
(d) MAX for A24.
Note that we used AO-THC for all calculations presented here.
The dotted lines in (a) and (b) indicate 1 kcal/mol.
\insertnew{A non-variational collapse was observed when there is a sudden jump in RMSD and/or MAX.}
}
\end{figure}

In \cref{fig:hf}, we assess the accuracy of THC-RI-K when used in HF calculations.
Unfortunately, THC-RI-K does not guarantee variationality in the energy evaluation 
so we observed striking non-variational failures for small $c_\text{ISDF}$.
Comparing W4-11 ((a) and (b)) and A24 ((c) and (d)) in \cref{fig:hf},
it is obvious that THC-RI-K works better in an absolute sense for non-covalent interaction energies.
\insertnew{This is important because of the small magnitude of the A24 energy differences.}
It is also important to note that the factorization becomes more inaccurate with increasing size of basis set.
In particular, for the largest basis set examined here (aug-cc-pVTZ)
THC-RI-K
exhibits a more inaccurate compression for a given $c_\text{ISDF}$ than other basis sets.
From rank considerations, it is not surprising that THC becomes more inaccurate with a larger basis set.
Based on (b) and (d), we conclude that $c_\text{ISDF} = 8.5$ 
approaches the accuracy of exact RI-K with an error less than 1 kcal/mol.
Furthermore, this value was large enough to have an overall numerically stable SCF calculation in both W4-11 and A24.

We further assess the utility of THC-RI-K in the context of an RSH functional, $\omega$B97X-V.
The exact exchange is now evaluated with 100\% of the long-range Coulomb operator, $\text{erf}(\omega r_{12})/r_{12}$
along with 16.7\% of the short-range Coulomb operator, $\text{erfc}(\omega r_{12})/r_{12}$ where the range-separation parameter $\omega$ is 0.3.
Since the short-range contribution is substantially reduced, it is possible to see
whether the accuracy of THC-RI-K varies for each component of the Coulomb operator.
\begin{figure}[h!]
\includegraphics[scale=0.85]{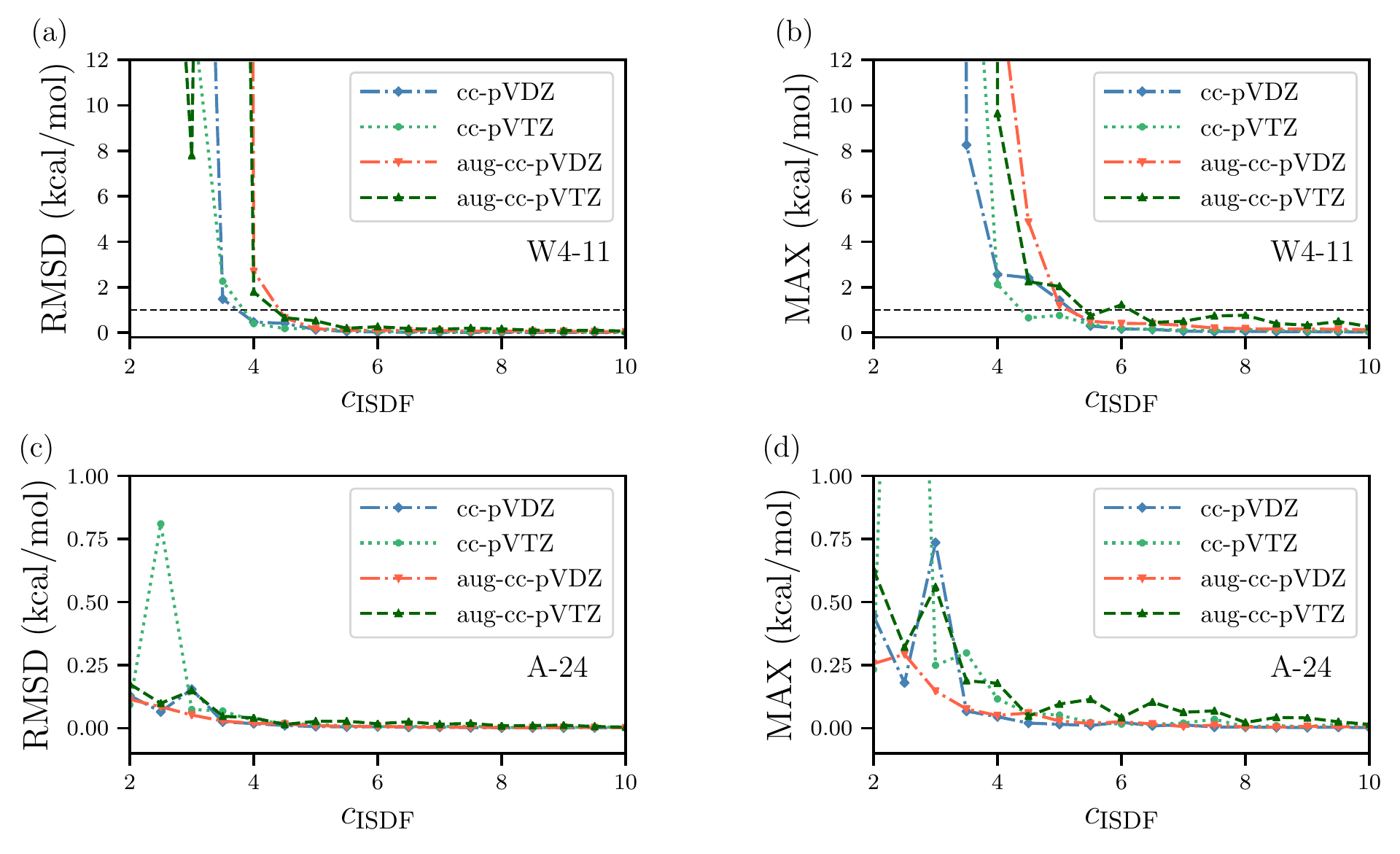}
\caption{\label{fig:wb97xv}
Same as \cref{fig:hf} except that the results here are for $\omega$B97X-V.
}
\end{figure}
The assessment of accuracy of $\omega$B97X-V with THC-RI-K is presented in \cref{fig:wb97xv}.
Similarly to HF, variational collapse was observed in some systems for small $c_\text{ISDF}$. 
With sufficiently large $c_\text{ISDF}$, such instability did not occur anymore.
We also note that 
it was easier to achieve numerical stability in THC-RI-K $\omega$B97X-V with a smaller $c_\text{ISDF}$ than
that of THC-RI-K HF. This indicates that the variational collapse may be caused by mainly the short-range component of exact exchange.
Furthermore, a faster convergence with respect to $c_\text{ISDF}$ is achieved with $\omega$B97X-V.
This is most obvious from comparing \cref{fig:hf} (b) and (d) with \cref{fig:wb97xv} (b) and (d).
This suggests that it is possible that long-range exact exchange may be easier to compress than the short-range one.
More thorough analysis for applying THC to range-separated Coulomb operators will be studied in the future.
Our recommendation for $c_\text{ISDF}$ in the case of $\omega$B97X-V is 6.5 for all basis sets considered in this work.

\subsection{THC-RI-MP2}
\begin{figure}[h!]
\includegraphics[scale=0.85]{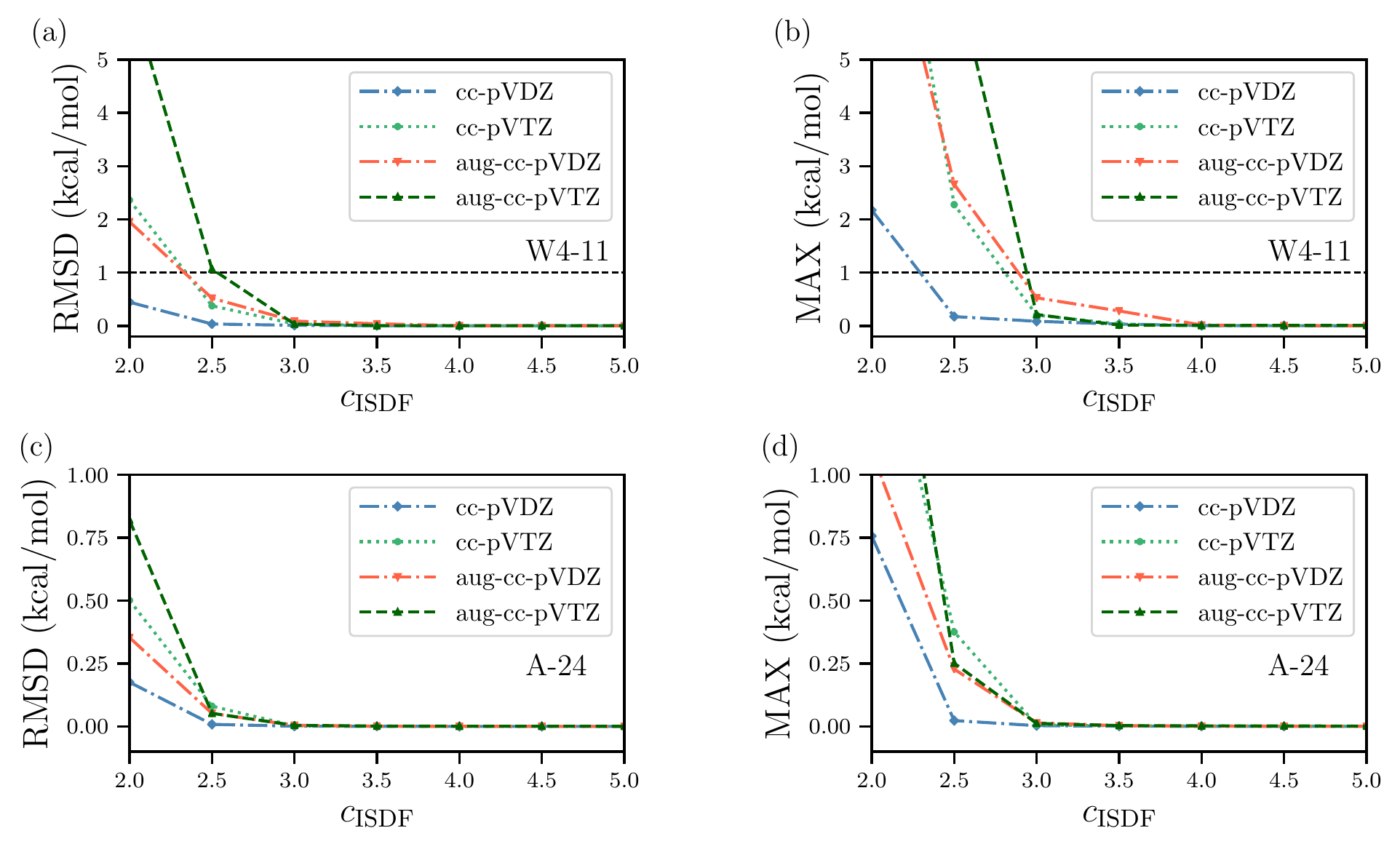}
\caption{\label{fig:mp2}
Same as \cref{fig:hf} except that the results here are for MP2.
Note that we used MO-THC (i.e., occupied-virtual block only) for all calculations presented here.
}
\end{figure}
THC-RI-MP2
is an easier example than THC-RI-K
because of the fact that
one can employ MO-THC
without causing numerical instability.
The resulting performance of THC-RI-MP2 is exceptionally accurate for both W4-11 and A24 sets as shown in \cref{fig:mp2}.
With $c_\text{ISDF} = 3.0$, based on RMSD THC-RI-MP2 is statistically indistinguishable
from RI-MP2. 
As before, we observe more difficulties in the W4-11 set than in the A24 set and also with larger basis sets.
Nevertheless, with $c_\text{ISDF} = 3.5$, regardless of the underlying AO basis sets
it is possible to achieve 1 kcal/mol accuracy (W4-11 MAX) as shown in (b) and $<$ 0.05 kcal/mol (A24 MAX) as shown in (d) of \cref{fig:mp2}.
We note that cc-pVDZ shows a faster convergence in both data sets compared to other larger basis sets.
This is consistent with studies by Mart{\'i}nez and co-workers where they focused on THC-RI-MP2 with cc-pVDZ \cite{Parrish2012,KokkilaSchumacher2015,Song2016}.

Based on these studies, we recommend $c_\text{ISDF} = 2.5$ for cc-pVDZ and $c_\text{ISDF} = 3.0$ for other larger basis sets for
THC-RI-MP2. THC-RI-MP2 achieves remarkably accurate factorization with only about three times more interpolation points than the number of auxiliary basis functions.
We emphasize that for cc-pVDZ we obtain a grid representation for THC-RI-MP2 as compact as
what was obtained from hand-optimized grids for cc-pVDZ developed by Mart{\'i}nez and co-workers \cite{KokkilaSchumacher2015}.
In particular, for water clusters, they obtained a total of 219 optimized grid points per water molecule that achieves more or less exact accuracy.
In our case, for water molecules included in the A24 set, 210 grid points per water molecule were enough to converge to RI-MP2. The real strength of ISDF point selection is that we do not have to hand-optimize the grid representation for each basis set
to achieve similar compactness.
The general procedure discussed in this paper can be combined with {\it any} available AO basis sets.
Given the timing benchmarks provided by Mart{\'i}nez's group \cite{KokkilaSchumacher2015}, we 
expect that THC-RI-MP2 combined with ISDF will be an accurate and efficient electronic structure tool even without considering any sparsity.

In passing, we note that with AO-THC we were not able to reach the same accuracy as MO-THC with the same $c_\text{ISDF}$.
Nonetheless, we were able to achieve chemical accuracy (1 kcal/mol) with $c_\text{ISDF} = 5.0$ for all basis sets considered in this work.
In the presence of diffuse basis functions, AO-THC RI-MP2 exhibits much slower convergence to RI-MP2 with respect to $c_\text{ISDF}$. We strongly recommend MO-THC over AO-THC for THC-RI-MP2 although nuclear gradient theory can be 
more complicated \cite{Song2017a}.

\subsection{THC-RI-MP3}
As mentioned in Section \ref{sec:mp3}, 
THC-RI-MP3 is more challenging than THC-RI-MP2 and THC-RI-HF. 
This is because it requires an accurate representation for all of the MO blocks in the two-electron tensor.
\begin{figure}[h!]
\includegraphics[scale=0.85]{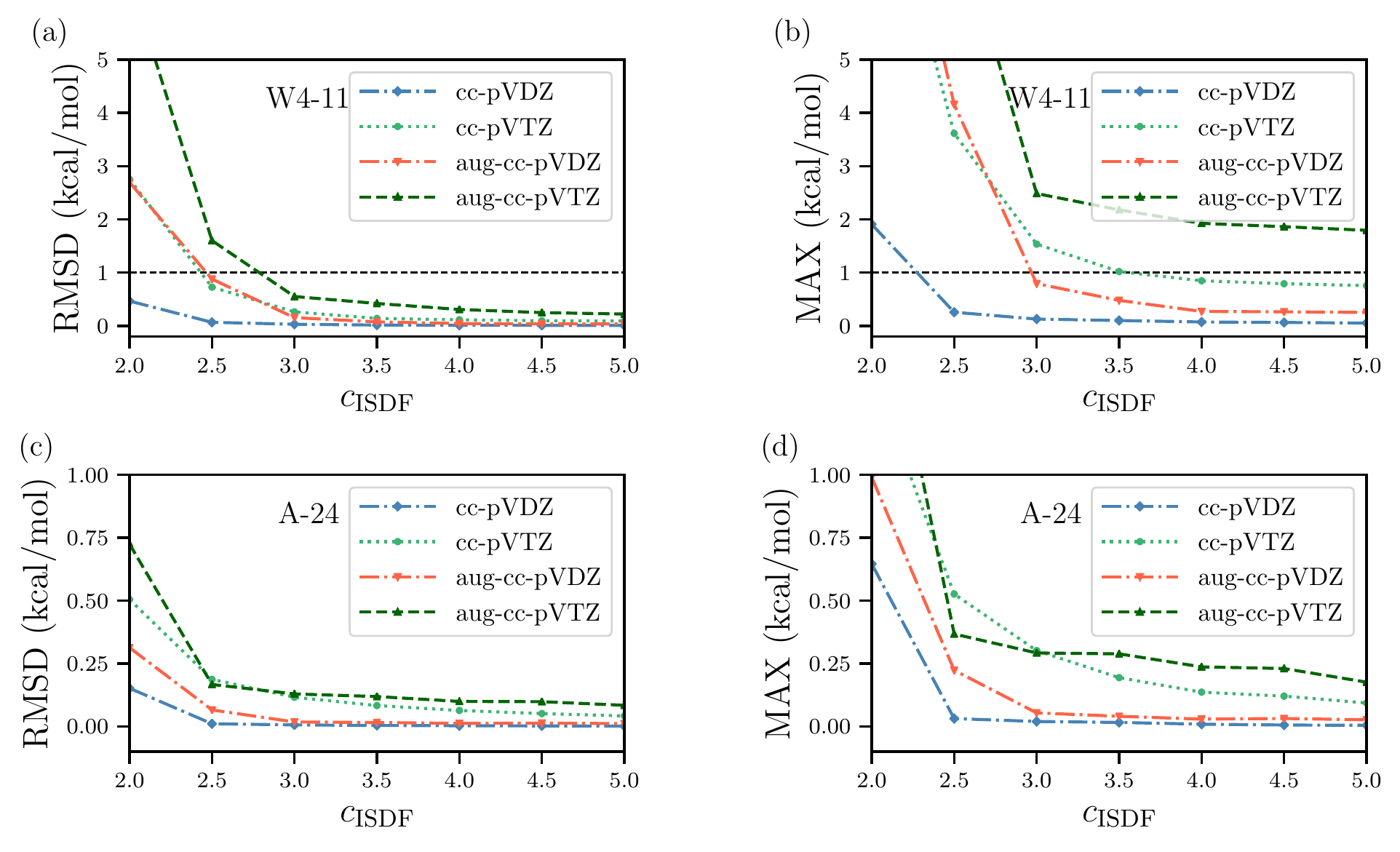}
\caption{\label{fig:mp3}
Same as \cref{fig:hf} except that the results here are for MP3.
Note that we used MO-THC for all calculations presented here. \insertrev{The slow convergence for large basis sets (cc-pVTZ and aug-cc-pVTZ) is likely due to the amplitude factorization as indicated by our investigation of THC-RI-MP2 with both factorized integrals and amplitude. See the Supporting Information for details.}
}
\end{figure}

In \cref{fig:mp3}, we present the THC-RI-MP3 results for W4-11 and A24 sets.
Based on RMSDs,  with $c_\text{ISDF} = 3.0$ THC-RI-MP3 invokes an error below 1 kcal/mol for all basis sets.
However, in its worst-performing chemical system of W4-11 (see \cref{fig:mp3} (b)),
for cc-pVTZ and aug-cc-pVTZ basis sets we observe remarkably slow convergence of the energy 
with respect to $c_\text{ISDF}$. 
This data point corresponds to the atomization energy of \ce{P4}.
We performed a THC-RI-MP3 calculation with $c_\text{ISDF} = 10.0$ for this single data point
and the error becomes 1.33 kcal/mol.
A similar slow convergence was observed in A24 for ethyne dimer although the error is not as large as that of \ce{P4}.
This is partly due to its much smaller energy scale (1.45 kcal/mol) compared to that of \ce{P4} (290.58 kcal/mol).

\insertrev{We found that the amplitude factorization defined in \cref{eq:tfacfinal} fails to faithfully represent amplitude
for larger basis sets. Since we cannot perform THC-RI-MP3 without factorized amplitude due to the steep computational cost, we assessed the accuracy of amplitude factorization by evaluating the THC-RI-MP2 with both factorized integrals and amplitudes. It turns out that the combination of integral and amplitude factorizations results in far worse energies particularly for larger basis sets. Interested readers are referred to the Supporting Information for further details (see Fig. S1). It appears that direct factorization of amplitude as proposed in ref. \citenum{schutski2017tensor} should be able to achieve greater accuracy at an increased cost via potentially difficult non-linear optimization.}

The slow convergence is also associated with the metric matrix $\tilde{\mathbf S}$ becoming more singular
with increasing size of basis set.
This singularity then affects the accuracy of the least-squares fit due to thresholding in the formation of the pseudoinverse.
This was not a serious problem for MP2 because the occupied-virtual metric is relatively better conditioned than that of the virtual-virtual block.
It turns out that the singularity problem is not as serious for the uniform grid as is done in planewave codes \cite{Hu2017,malone2018overcoming}.
It will be interesting to explore a mixed grid representation between Becke's and uniform grids to
see if it helps to cope with the numerical instability.

Given these results, we recommend
$c_\text{ISDF} = 2.5$ for cc-pVDZ and $c_\text{ISDF} = 4.0$ for aug-cc-pVDZ.
For the larger basis sets, cc-pVTZ and aug-cc-pVTZ, we recommend $c_\text{ISDF} = 5.0$
based on statistical performance. 
However, use of THC-RI-MP3 with these basis sets requires caution due to the remarkably slow convergence rate
observed in some chemical systems such as \ce{P4}.

\subsection{Summary of Recommendations}\label{sec:summary}
We summarize the recommendations made for each method throughout the previous sections in Table \ref{tab:summary}.
\begin{table}[h]
\begin{tabular}{|c|c|c|c|c|}\hline
Basis Set & HF & $\omega$B97X-V & MP2 & MP3 \\ \hline
cc-pVDZ &8.5 &6.5&2.5& 2.5\\ \hline
cc-pVTZ &8.5&6.5&3.0& 5.0\\ \hline
aug-cc-pVDZ &8.5&6.5&3.0& 4.0\\ \hline
aug-cc-pVTZ &8.5&6.5&3.0&$>$5.0\\ \hline
\end{tabular}
\caption {Summary of recommendations for $c_\text{ISDF}$ in THC-RI to achieve the same accuracy as various exact RI methods.
For HF and $\omega$B97X-V, AO-THC was employed while MO-THC was employed for MP2 and MP3.
}
\label{tab:summary}
\end{table}
The most successful application of THC-RI combined with the ISDF point selection was THC-RI-MP2.
We shall further provide practical information for THC-RI methods.
In particular, we present $N_\text{AO}$, $N_\text{X}$, and $N_\text{IP}$ (of each THC-RI method) of the oxygen atom (Table \ref{tab:oxygen}) and the hydrogen atom (Table \ref{tab:hydrogen}) 
for the basis sets considered here. 
\begin{table}[h]
\begin{tabular}{|c|c|c|c|c|c|c|}
\cline{4-7}
\multicolumn{3}{c}\: & \multicolumn{4}{|c|}{$N_\text{IP}$}\\\hline
Basis Set & $N_\text{AO}$ & $N_\text{X}$ &
HF
& $\omega$B97X-V
& MP2 
& MP3
\\ \hline
cc-pVDZ & 14 & 70 & 593$^*$ & 455 & 175 & 175  \\ \hline
cc-pVTZ & 30 & 81 &684$^*$ &525$^*$ & 243& 405 \\ \hline
aug-cc-pVDZ & 23 & 86 & 730$^*$ & 559 & 258 &  344\\ \hline
aug-cc-pVTZ & 46 & 106 & 898$^*$ & 688$^{*}$& 318& 530\\ \hline
\multicolumn{7}{l}{
$^{*}$ Some redundancies in the CVT process were removed so that we have fewer points}\\
\multicolumn{7}{l}{than $c_\text{ISDF} \times N_\text{X}$.}
\end{tabular}
\caption {
The number of AO basis functions, auxiliary basis functions, and (recommended) interpolation points for the oxygen atom for performing THC-RI-MP2 with various basis sets.
}
\label{tab:oxygen}
\end{table}

\begin{table}[h]
\begin{tabular}{|c|c|c|c|c|c|c|}
\cline{4-7}
\multicolumn{3}{c}\: & \multicolumn{4}{|c|}{$N_\text{IP}$}\\\hline
Basis Set & $N_\text{AO}$ & $N_\text{X}$ &
HF
& $\omega$B97X-V
& MP2 
& MP3
\\ \hline
cc-pVDZ & 5 & 14 &119&91& 35& 35 \\ \hline
cc-pVTZ & 14 & 30 & 255 &195 & 90& 150 \\ \hline
aug-cc-pVDZ & 9 & 23 & 195 & 149& 69& 92 \\ \hline
aug-cc-pVTZ & 23 & 46 & 391 & 299 & 138 & 230\\ \hline
\end{tabular}
\caption {
Same as Table \ref{tab:oxygen} but for the hydrogen atom.
}
\label{tab:hydrogen}
\end{table}

\insertrev{The molecules included in W4-11 and A24 are relatively small in size so it may be unclear 
how much compression one is achieving with those recommended values. We provide $N_\text{AO}$, $N_\text{X}$, $n_\text{occ}$, and $N_\text{IP}$ for $c_\text{ISDF}\in[2.0, 10.0]$ with an increment of 0.5. Interested readers are referred to the Supporting Information for these parameters. We shall briefly discuss those parameters for the largest system in each of W4-11 and A24 sets.
The largest molecule in W4-11 is propane (\cref{tab:propane}) and the one in A24 is ethene dimer (\cref{tab:ethene}). 
As shown in these tables, except for the case of MP2/cc-pVDZ for propane, THC always achieves compression for the pertinent integrals even for these small molecules. In the next section, we present more numerical data for larger systems to demonstrate the power of THC combined with ISDF grids further.
\begin{table}[h]
\begin{tabular}{|c|r|r|r|r|r|r|r|r|r|r|}\cline{8-11}
\multicolumn{7}{c}\: & \multicolumn{4}{|c|}{$N_\text{IP}$}\\\hline
Basis & $N_\text{AO}$ & $N_\text{X}$ & $n_\text{occ}$ & $n_\text{occ} n_\text{vir}$	 & $N_\text{AO}^2$ &$N_\text{SP}$& HF & $\omega$B97X-V & MP2 & MP3\\ \hline
cc-pVDZ & 82 & 394 & 69 & 897 & 6724 & 902 & 3339 & 2559 & 984 & 984 \\\hline
cc-pVTZ & 202 & 483 & 189 & 2457 & 40804 & 3061 & 4090 & 3134 & 1445 & 2410 \\\hline
aug-cc-pVDZ & 141 & 514 & 128 & 1664 & 19881 & 2278 & 4354 & 3333 & 1537 & 2053 \\\hline
aug-cc-pVTZ & 322 & 686 & 309 & 4017 & 103684 & 6542 & 5816 & 4451 & 2058 & 3430 \\\hline
\end{tabular}
\caption {
$N_\text{AO}$, $N_\text{X}$, $n_\text{occ}$, $n_\text{occ}n_\text{vir}$, $N_\text{AO}^2$, and $N_\text{IP}$ of various methods for propane in W4-11. Note that for $N_\text{IP}$ we present the corresponding recommended values for each method.
}
\label{tab:propane}
\end{table}
\begin{table}[h]
\begin{tabular}{|c|r|r|r|r|r|r|r|r|r|r|}\cline{8-11}
\multicolumn{7}{c}\: & \multicolumn{4}{|c|}{$N_\text{IP}$}\\\hline
Basis & $N_\text{AO}$ & $N_\text{X}$ & $n_\text{occ}$ & $n_\text{occ} n_\text{vir}$	 & $N_\text{AO}^2$ &$N_\text{SP}$& HF & $\omega$B97X-V & MP2 & MP3\\ \hline
cc-pVDZ & 96 & 336 & 80 & 1280 & 9216 & 1152 & 2848 & 2176 & 832 & 832 \\ \hline
cc-pVTZ & 232 & 564 & 216 & 3456 & 53824 & 3642 & 4776 & 3664 & 1688 & 2816 \\ \hline
aug-cc-pVDZ & 164 & 472 & 148 & 2368 & 26896 & 2920 & 4007 & 3064 & 1408 & 1880 \\ \hline
aug-cc-pVTZ & 368 & 792 & 352 & 5632 & 135424 & 8062 & 6711 & 5136 & 2372 & 3956 \\ \hline
\end{tabular}
\caption {
Same as \cref{tab:propane}, but for ethene dimer in A-24.
}
\label{tab:ethene}
\end{table}
}

\subsection{\insertrev{Application to Larger Systems: \ce{(H2O)_{$N$}} and \ce{C20}}}\label{sec:larger}
\insertrev{With the values determined through benchmarking over the W4-11 and A24 sets,
we applied $\omega$B97X-V with THC-RI-K and THC-RI-MP2 to larger systems than
those in W4-11 and A24. The basis set used here is cc-pVTZ which should be a good compromise between accuracy and cost for DFT and MP2 calculations.}

\begin{table}[h]
\begin{tabular}{|c|r|}\hline
$N$ & Error/$N$ \\ \hline
4 & 12.6 \\ \hline
8 & 9.2 \\ \hline
12 & 14.8 \\ \hline
16 & 2.7 \\ \hline
20 & 19.7 \\ \hline
\end{tabular}
\caption {
Error per water molecule ($\mu$$E_h$) in the absolute energies of THC-RI-K $\omega$B97X-V compared to RI-K $\omega$B97X-V for \ce{(H2O)_$N$}.
The cc-pVTZ basis set and its corresponding auxiliary basis set are used and $c_\text{ISDF} = 6.5$ as determined in \cref{sec:summary}.
}
\label{tab:h2orik}
\end{table}
\begin{table}[h]
\begin{tabular}{|c|r|}\hline
$N$ & Error/$N$ \\ \hline
8 & 22.8 \\ \hline
16 & 25.2 \\ \hline
32 & 29.8 \\ \hline
\end{tabular}
\caption {
Error per water molecule ($\mu$$E_h$) in the absolute energies of THC-RI-MP2 compared to RI-MP2 for \ce{(H2O)_$N$}.
The cc-pVTZ basis set and its corresponding auxiliary basis set are used and $c_\text{ISDF} = 3.0$ as determined in \cref{sec:summary}.
}
\label{tab:h2omp2}
\end{table}
\insertrev{First, we applied THC to water clusters ranging from $N=2$ to $N=20$ for $\omega$B97X-V and $N=8, 16, 32$ for MP2.
{The geometries for $N=8, 16$ were obtained by truncating geometries reported in ErgoSCF\cite{ergoscf}. Other geometries were taken from the Cambridge Cluster Database\cite{camb}.}
It is shown in \cref{tab:h2orik} and \cref{tab:h2omp2} that when growing the number of water molecules, the error per water molecule does not increase significantly. They are on the order of 10 - 30 $\mu$$E_h$ which provides enough margin for error cancellation to achieve chemical accuracy.}

\insertrev{Next, we also apply these methods to compute the relative energetics of four different Jahn-Teller distorted cage structures of \ce{C20} (C$_\text{2h}$, D$_\text{2h}$, C$_\text{i}$, and D$_\text{3h}$) taken from Manna and Martin's work\cite{Manna2016}.
As noted in refs. \citenum{Manna2016} and \citenum{lee2019distinguishing}, those four structures differ only slightly in energy. In fact, 
at the level of MP2, there is near-degeneracy between C$_\text{i}$ and D$_\text{3h}$ cages and also between
C$_\text{2h}$ and D$_\text{2h}$ cages. Achieving accurate integral factorization consistently over all four structures is crucial to capture this qualitative feature.} 
\begin{table}[h]
\begin{tabular}{|c|r|r|}\hline
Structure & THC-RI-K & THC-RI-MP2 \\ \hline
C$_\text{2h}$ & 10.95 & 9.54 \\ \hline
D$_\text{2h}$ & 11.35 & 18.94 \\ \hline
C$_\text{i}$ & 12.88 & 20.02 \\ \hline
D$_\text{3h}$ & 14.33 & 61.05 \\ \hline
\end{tabular}
\caption {
Error per atom ($\mu$$E_h$) in the absolute energies of THC-RI-K $\omega$B97X-V and THC-RI-MP2 compared to corresponding reference methods for different structures of \ce{C20}.
The cc-pVTZ basis set and its corresponding auxiliary basis set are used and $c_\text{ISDF} = 6.5$ for THC-RI-K and $c_\text{ISDF}=3.0$ for THC-RI-MP2 as determined in \cref{sec:summary}.
}
\label{tab:c20}
\end{table}
\begin{table}[h]
\begin{tabular}{|c|r|r|r|r|}\hline
Structure & RI-K & THC-RI-K & RI-MP2 & THC-RI-MP2 \\ \hline
C$_\text{2h}$ & 0.17 & 0.14 & 0.83 & 0.70 \\ \hline
D$_\text{2h}$ & 0.02 & 0.00 & 0.83 & 0.82 \\ \hline
C$_\text{i}$ & 0.00 & 0.00 & 0.00 & 0.00 \\ \hline
D$_\text{3h}$ & 0.22 & 0.24 & 0.01 & 0.52 \\ \hline
\end{tabular}
\caption {
Relative energies (kcal/mol) of four cage structures of \ce{C20} obtained from RI-K $\omega$B97X-V, THC-RI-K $\omega$B97X-V, RI-MP2, and THC-RI-MP2.
The cc-pVTZ basis set and its corresponding auxiliary basis set are used and $c_\text{ISDF} = 6.5$ for THC-RI-K and $c_\text{ISDF}=3.0$ for THC-RI-MP2 as determined in \cref{sec:summary}.
}
\label{tab:c202}
\end{table}
\insertrev{In \cref{tab:c20}, we present the absolute energy error per atom for THC-RI-K and THC-RI-MP2. 
THC-RI-K shows 10-15 $\mu$$E_h$ per atom errors whereas THC-RI-MP2 error can be as large as 61 $\mu$$E_h$ per atom.
This yields the relative energies in \cref{tab:c20}. Both RI and THC-RI predict C$_\text{i}$ to be the lowest structure. The error of THC-RI-K ranges from 0.02 kcal/mol to 0.03 kcal/mol which is essentially chemically indistinguishable from RI-K. On the other hand, THC-RI-MP2 does not capture the near-degeneracy between C$_\text{i}$ and D$_\text{3h}$. Nevertheless, the difference here is on the order of 0.5 kcal/mol. }

\insertrev{Based on these larger applications, we conclude that the numerically observed error caused by THC combined with ISDF grid points grows only linearly with system size and can achieve chemical accuracy for larger systems such as \ce{C20}. This suggests that one can set $c_\text{ISDF}$ independent of system size to achieve a fixed accuracy per atom or molecule.}
\subsection{Estimated Crossovers}

\begin{table}[h]
\begin{tabular}{|c|c|c|c|c|}\hline
$N_e$ & HF & $\omega$B97X-V & MP2 & MP3\\\hline
16 & 0.08 & 0.14 & 0.01 & 0.00 \\ \hline
32 & 0.16 & 0.27 & 0.02 & 0.00 \\ \hline
64 & 0.32 & 0.55 & 0.04 & 0.01 \\ \hline
128 & 0.65 & 1.10 & 0.09 & 0.04 \\ \hline
256 & 1.29 & 2.20 & 0.17 & 0.17 \\ \hline
512 & 2.58 & 4.40 & 0.34 & 0.69 \\ \hline
1024 & 5.16 & 8.79 & 0.68 & 2.76 \\ \hline
2048 & 10.32 & 17.58 & 1.36 & 11.05 \\ \hline
4096 & 20.65 & 35.16 & 2.73 & 44.19 \\ \hline
8192 & 41.30 & 70.32 & 5.45 & 176.75 \\ \hline
16384 & 82.59 & 140.64 & 10.90 & 706.99 \\ \hline
\end{tabular}
\caption {
Estimated cost ratio between conventional RI methods and THC-RI methods for a range of number of electrons ($N_e$)
assuming stoichiometry of (\ce{H2O}$)_n$ and $c_\text{ISDF}$ in Table \ref{tab:oxygen} and Table \ref{tab:hydrogen}.
If the ratio is greater than 1, then the THC algorithm is faster than the RI algorithm and vice versa.
Note that these are based on computational bottlenecks of each algorithm and no special treatment for sparsity was assumed.
\insertnew{Non-trivial prefactors associated with each method were neglected or simplified. For instance, the cost of THC-RI-MP3 was assumed to be $\mathcal O(16 N_\text{IP}^4)$ whereas THC-RI-HF was simply assumed to be $\mathcal O(N_\text{AO}N_\text{IP}^2+n_\text{occ}N_\text{IP}^2)$.
Similarly, we used
$\mathcal O(7\times n_\text{occ}n_\text{vir}N_\text{IP}^2)$ for THC-RI-MP2, 
$\mathcal O(n_\text{occ}^2n_\text{vir}^2N_\text{X})$ for RI-MP2, 
$\mathcal O(N_\text{AO}^2n_\text{occ}N_\text{X})$ for RI-K,
and 
$\mathcal O(n_\text{occ}^2n_\text{vir}^4
+ n_\text{occ}^4n_\text{vir}^2
+ 4\times n_\text{occ}^3n_\text{vir}^3
)$ for RI-MP3.
}
}
\label{tab:crossover}
\end{table}
\insertnew{
RI methods are well-known to be effective for molecules up to
moderately large in size, yielding 
shorter times to solution
than 
conventional AO quantum chemistry, 
despite poorer formal scaling for HF and DFT exact exchange.
THC methods, as we have explored in detail, 
reduce the formal scaling at the cost of quite large prefactors.
It is then fairly straightforward
to estimate rough crossovers
for HF exchange, exact exchange in DFT ($\omega$B97X-V),
MP2 and MP3.
These estimates for the medium-sized cc-pVTZ basis
are computed in Table \ref{tab:crossover} based on Table \ref{tab:oxygen} and Table \ref{tab:hydrogen}, using the
assumption of (H${}_2$O)$_n$ stoichiometry, the recommended
$c_\text{ISDF}$ values already given, 
and operation counts for the rate-determining steps via RI and THC-RI.
The main point is that in the absence of any special treatment for sparsity
the crossovers are on the order of
100--200 electrons for exchange,
1000--2000 electrons for MP2,
and
500--1000 electrons for MP3,
which are large but not unreachable, especially for exchange and MP2.
}

\begin{figure}[h!]
\includegraphics[scale=0.85]{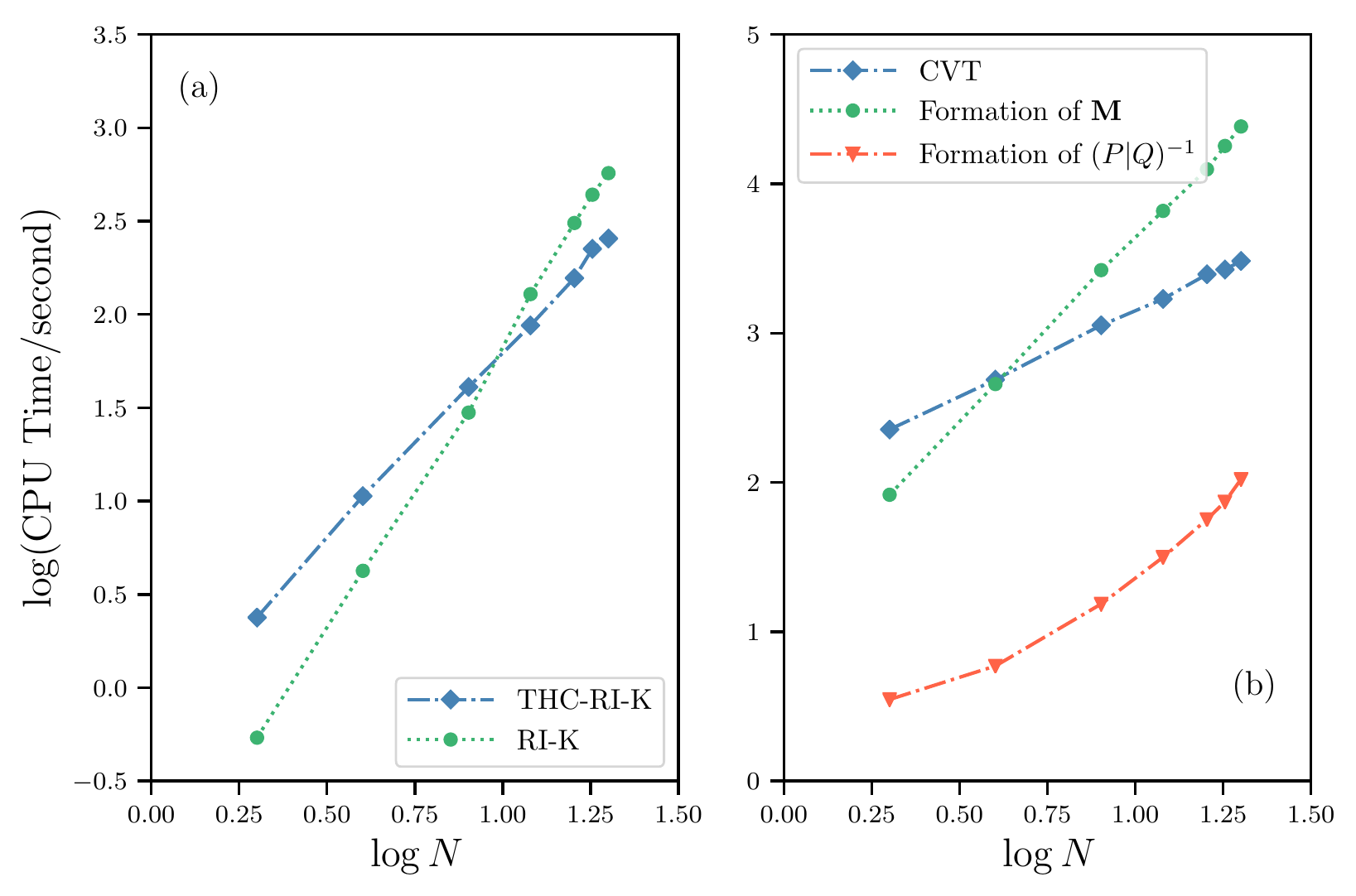}
\caption{\label{fig:timing}
Measured CPU timing for RI-K, THC-RI-K and preparation steps for THC-RI-K in the case of water clusters ($N=2, 4, 8, 12, 16, 18, 20$) with $\omega$B97X-V/cc-pVTZ and $c_\text{ISDF}=6.5$. 
(a) Comparison of RI-K and THC-RI-K and
(b) CPU times for CVT, the formation of $\mathbf M$, and the formation of $(P|Q)^{-1}$.
}
\end{figure}

\insertnew{
As an example, we report CPU timing for running $\omega$B97X-V on water clusters ($N=2, 4, 8, 12, 16, 18, 20$) as shown in \cref{fig:timing}.
\insertrev{We used 32 cores on an AMD Opteron 6376 Processor to obtain the timing results.}
\insertrev{Even with our pilot implementation (i.e., no sparsity beyond the shell-pair sparsity)}, 
the crossover between THC-RI-K and RI-K occurs at about 10 water molecules ($N_e=80$).
As shown in \cref{fig:timing} (b), however, the preparation steps for obtaining the THC factorization 
takes most of the time for systems considered in this timing benchmark.
Asymptotically, since the preparation steps are cubic scaling and occur only once at the beginning of the SCF,
there will be eventually a crossover between THC-RI-K and RI-K including preparation steps for each.
\insertrev{Given this promising result in \cref{fig:timing}, it will be interesting to further optimize computational kernels and seek earlier crossovers.}
}
\section {Conclusions and Outlook}
In this work, we investigated the utility of tensor hypercontraction (THC) combined with the resolution-of-the-identity (RI) approximation and the interpolative separable density fitting (ISDF) grid point selection.
We presented a modified ISDF point selection that can be straightforwardly combined with
\insertrev{Becke-type} atom-centered grids typically used in density functional theory calculations.
Within this framework, we assessed the accuracy of THC-RI with ISDF points
when applied to
the evaluation of exact exchange for Hartree-Fock (HF) and a range-separated hybrid (RSH) functional $\omega$B97X-V
and
the second and the third-order M{\o}ller-Plesset perturbation theory (MP2 and MP3) correlation energy.
Since the framework developed in this work is general, we were able to investigate
the accuracy of these methods with 
multiple basis sets (cc-pVDZ, cc-pVTZ, aug-cc-pVDZ and aug-cc-pVDZ) 
over 140 data points for total atomization energy included in the W4-11 set \cite{Karton2011}
and 24 data points for non-covalent interaction energy included in the A24 set \cite{Rezac2013}.
For each basis set and method, we made a recommendation for $c_\text{ISDF}$, which is a single parameter
that is used in the ISDF point selection. The number of interpolation points $N_\text{IP}$ is determined by $
N_\text{IP} = c_\text{ISDF}\times N_\text{X}$ where $N_\text{X}$ is the number of RI basis functions.

{\it Exact exchange:} A cubic-scaling algorithm for exact exchange (K) was formulated within the THC-RI framework. 
In THC-RI-K, fitting molecular orbital integrals (i.e., MO-THC) directly leads to a modified Fock matrix.
Therefore, for simplicity, we investigated THC with fitting atomic orbital integrals (i.e., AO-THC).
Since THC-RI-K does not guarantee variationality, subsequent SCF calculations with it often exhibits 
non-variational collapse. 
Regardless of the basis set size, we recommend $c_\text{ISDF} = 8.5$ and $c_\text{ISDF} = 6.5$ for THC-RI-K when used in HF and $\omega$B97X-V, respectively. This was found to be sufficient in achieving chemical accuracy (1 kcal/mol) for every data point in the W4-11 benchmarks considered in this work (and much better for the A24 intermolecular interactions).

{\it MP2:} A quartic-scaling algorithm for evaluating MP2 correlation energy was formulated, which is identical to what was proposed by Mart{\'i}nez and co-workers. Using MO-THC, we achieved more or less the same accuracy as RI-MP2 with $c_\text{ISDF} = 2.5$ for cc-pVDZ and $c_\text{ISDF} = 3.0$ for cc-pVTZ, aug-cc-pVDZ and aug-cc-pVTZ.
Such accuracy was also achieved with hand-optimized grids developed for each basis set by Mart{\'i}nez and co-workers \cite{KokkilaSchumacher2015}. We emphasize that the ISDF point selection can achieve both compactness and accuracy of
the hand-optimized grids and is more general because it can be straightforwardly applied to any basis sets that a user specifies.

{\it MP3:} We developed a new quartic-scaling algorithm for MP3 that utilizes THC-factorized first-order wavefunction amplitudes. With factorized amplitudes, the THC-RI-MP3 algorithm no longer has the double Laplace transformation (LT) in it. Instead, the transformation is folded into the factorized amplitudes.
As a result, this algorithm removes a prefactor of 49 (assuming 7 quadrature points for LT) in the previously developed THC-RI-MP3 algorithm \cite{Hohenstein2012}.
The assessment of this algorithm with MO-THC over W4-11 and A24
suggested that
$c_\text{ISDF} = 2.5$ for cc-pVDZ, $c_\text{ISDF} = 4.0$ for aug-cc-pVDZ, and $c_\text{ISDF} > 5.0$ for cc-pVTZ and aug-cc-pVTZ basis sets.
Since MP3 needs all MO blocks of two-electron integrals, it is more difficult for THC-RI to achieve high accuracy than it is for MP2.
This was attributed to the metric matrix (particularly for the virtual-virtual block)  becoming more singular with larger basis sets.
\insertrev{Furthermore, the amplitude factorization through a simple least-squares fit was found to be insufficient for accurate results in large basis sets (aug-cc-pVTZ and cc-pVTZ).}

\insertrev{{\it Stress tests on larger systems:} 
We applied THC-RI-K $\omega$B97X-V and THC-RI-MP2 to larger systems such as \ce{(H2O)_$N$} and \ce{C20}. This was done by using cc-pVTZ combined with ISDF grid points generated from recommended $c_\text{ISDF}$ values for this basis set.
Increasing the size of water cluster, we concluded that the THC error scales linearly with system size. Based on the results from \ce{(H2O)_$N$} and \ce{C20}, we concluded that the THC error is about 20-60 $\mu$$E_h$. When combined with RI (whose error is about 50 $\mu$$E_h$ per atom or molecule), this provides enough margin for error cancelation to achieve chemical accuracy. 
}

{\it Remaining challenges and outlook:}
Based on the ISDF point selection, it is clear that THC-RI-MP2 is easiest to make robust and accurate compared to other THC methods examined in this work.
Although the energy evaluation presented in this work is quite straightforward to implement, 
its nuclear gradient can be challenging to implement within the MO-THC formalism. 
For exact exchange, more thorough investigation to understand frequent variational collapse is needed and
further improvements are highly desirable. It may be interesting to investigate Dunlap's variational ansatz \cite{Dunlap2000a}
with the THC factorization to have a robust, variational fit.
\insertrev{For MP3 (and other correlated methods such as coupled-cluster theory),
there is a need to develop a more robust alternative fitting or representation for amplitude factorization.} 
Furthermore, it was difficult to observe any speed-up from THC-RI-MP3 compared to RI-MP3 for the systems studied 
in this work (i.e., the crossover is not early).
Therefore, it is important to implement THC-RI-MP3 with careful consideration of sparsity.
Since a majority of THC-RI algorithms work in 3-space (where locality plays a crucial role), sparsity \revrep{will}{may} greatly reduce the cost \insertrev{at the expense of more difficult implementation}.
Therefore, it is highly desirable to develop a sparsity-aware THC implementation as was done for scaled opposite spin MP2 \cite{Song2016}.
It will be interesting to see whether a sparse implementation of the THC-RI-MP3 algorithm provides a substantial reduction in its cost over the conventional RI-MP3 algorithm even for medium-sized molecules.

The application of THC-RI to other methods is worthwhile such as mean-field excited state methods (e.g., configuration interaction with singles and time-dependent density functional theory), more sophisticated random phase approximations such as second-order screened exchange \cite{gruneis2009making}, coupled cluster methods \cite{Hohenstein2012a}, orbital-optimized MP2 \cite{lee2018regularized}, and etc.
Furthermore, a theoretical question about whether short-range or long-range is easier to compress via THC will be interesting
to investigate. We are currently exploring some of these \insertrev{issues} in our group.

\section{Acknowledgement}
We thank Fionn Malone and Miguel Morales for pointing out the Coulomb singularity problem in THC for molecular systems in the beginning of this project. Furthermore, we thank Fionn Malone and Luke Bertels for useful discussions on the W4-11 and A24 benchmarks,
Evgeny Epifanovsky for help on verifying the Becke quadrature implementation used in this work,
and
Srimukh Prasad for providing water cluster geometries used for timing benchmarks as well as discussions related to them.
J. L. thanks Soojin Lee for consistent encouragement.
\insertnew{This material is based upon work supported by the U.S. Department of Energy, Office of Science, Office of Advanced Scientific Computing Research, Scientific Discovery through Advanced Computing (SciDAC) program.}

\bibliography{isdf_ms}
\bibliographystyle{achemso}
\end{document}